\documentclass[twocolumn,twocolappendix]{aastex631}

\def\lapp{\ifmmode\stackrel{<}{_{\sim}}\else$\stackrel{<}{_{\sim}}$\fi}
\def\gapp{\ifmmode\stackrel{>}{_{\sim}}\else$\stackrel{>}{_{\sim}}$\fi}
\usepackage{graphicx}
\usepackage{multirow}
\usepackage{color}
\usepackage{amsmath}
\usepackage{soul}
\usepackage{hyperref}
\newcommand{\fluxcgs}{\ensuremath{\mathrm{erg}\,\mathrm{s}^{-1}\,\mathrm{cm}^{-2}}}
\newcommand{\kms}{\ensuremath{\mathrm{km}\,\mathrm{s}^{-1}}}

\shorttitle{}
\shortauthors{}
\begin{document}

\title{Modeling X-ray and gamma-ray emission from redback pulsar binaries}

\correspondingauthor{Hongjun An}
\email{hjan@cbnu.ac.kr}

\author[0009-0004-9621-0252]{Minju Sim}
\author[0000-0002-6389-9012]{Hongjun An}
\affiliation{Department of Astronomy and Space Science, Chungbuk National University, Cheongju, 28644, Republic of Korea}
\author[0000-0003-2714-0487]{Zorawar Wadiasingh}
\affiliation{Department of Astronomy, University of Maryland, College Park, Maryland 20742, USA}
\affiliation{Astrophysics Science Division, NASA Goddard Space Flight Center,Greenbelt, MD 20771, USA.}
\affiliation{Center for Research and Exploration in Space Science and Technology, NASA/GSFC, Greenbelt, Maryland 20771, USA}

\begin{abstract}
We investigated the multiband emission from
the pulsar binaries XSS~J12270$-$4859, PSR~J2039$-$5617, and PSR~J2339$-$0533,
which exhibit orbital modulation in the X-ray and gamma-ray bands.
We constructed the sources' broadband spectral energy distributions
and multiband orbital light curves by supplementing our X-ray measurements
with published gamma-ray results, and we modeled the data using
intra-binary shock (IBS) scenarios.
While the X-ray data were well explained by synchrotron emission from electrons/positrons
in the IBS, the gamma-ray data were difficult to explain with
the IBS components alone. Therefore, we explored other scenarios
that had been suggested for gamma-ray emission from pulsar binaries:
(1) inverse-Compton emission in the upstream unshocked wind zone and
(2) synchrotron radiation from electrons/positrons interacting with a kilogauss magnetic field of the companion.
Scenario~(1) requires that the bulk motion of the wind 
substantially decelerates to $\sim$1000\,\kms\ before reaching the IBS for increased residence time, in which case
formation of a strong shock is untenable, inconsistent with the X-ray phenomenology.
Scenario~(2) can explain the data if we assume the presence of electrons/positrons with a Lorentz factor of $\sim10^8$ ($\sim 0.1$\,PeV)
that pass through the IBS and tap a substantial portion of the pulsar voltage drop.
These findings raise the possibility
that the orbitally-modulating gamma-ray signals from pulsar binaries can provide insights into
the flow structure and energy conversion within pulsar winds and particle acceleration nearing PeV energies in pulsars.
These signals may also yield greater understanding of kilogauss magnetic fields potentially hosted by the low-mass stars in these systems.

\end{abstract}

\bigskip
\section{Introduction}
\label{sec:intro}
Pulsar binaries, composed of a millisecond pulsar and a low-mass ($<M_\odot$) companion \citep[][]{Fruchter1988},
are thought to form when past accretion from the companion had spun up the pulsar over a timescale
of $\sim$Gyr \citep[][]{acrs82}. In pulsar binary systems, the pulsar is in a tight orbit
with a companion whose spin is tidally locked with the $<$\,day orbit.
The pulsar heats the pulsar-facing side of the companion, driving a strong wind from it.
Additionally, the companion may have a strong magnetic field ($B$) of $\ge$\,kG \citep[e.g.,][]{Sanchez2017,Kansabanik2021}.
The wind-wind or wind-$B$ interaction \citep[e.g.,][]{Harding1990,Wadiasingh2018}
between the pulsar (wind) and the companion (wind or $B$) can lead to shocks at their interface.
The shocks wrap around the pulsar if the companion's pressure/wind is stronger than the pulsar's, or vice versa.

The hallmark properties of emission from pulsar binaries
are multiband orbital modulations. The companion's blackbody (BB) emission orbitally modulates
due to pulsar heating and the ellipsoidal deformation of the star, exhibiting
characteristic day-night cycles \citep[e.g.,][]{Breton2013,Strader2019}.
Modeling these day-night cycles has proven useful for determining
system parameters such as orbital inclination and has facilitated estimations of the masses
of the neutron stars in pulsar binaries \citep[e.g.,][]{vanKerkwijk2011,Linares2020}.
Nonthermal X-ray emission, originating from the intra-binary shock (IBS), also shows
orbital modulation with a single- or double-peak structure \citep[e.g.,][]{Huang+12}
caused by Doppler beaming (relativistic aberration) by the IBS bulk flow.
This modulation, when combined with orbital parameters,
aids in probing particle acceleration and flow in the shock \citep[e.g.,][]{kra19,Merwe+2020,Cortes2022}.

Gamma-ray emission from pulsar binaries is dominated by pulsed magnetospheric radiation from the pulsar.
Other than a sharp dip in the light curves (LCs) of eclipsing systems \citep[][]{Corbet2022,Clark+2023},
the gamma-ray emission has been thought to be orbitally constant.
However, the Fermi Large Area Telescope \citep[LAT;][]{Atwood2009} has
discovered orbital modulation in the GeV band in a few pulsar binaries \citep[e.g.,][]{ntsl+18,ark18,Clark+2021},
suggesting that there should be other physical mechanisms for gamma-ray production
in these sources.
The GeV modulation observed in the `black widow'
\citep[BW; $<0.1M_{\odot}$ companion; see][for a list of BWs]{Swihart2022}
PSR~J1311$-$3430 was interpreted as
synchrotron radiation from the IBS particles \citep[][]{arjk+17,Merwe+2020} because its gamma-ray LC
peak appears at the phase where the IBS tail should be in the direction of
the observer's line of sight (LoS).
However, this synchrotron interpretation remains inconclusive
due to the absence of evidence for IBS X-rays (i.e., an orbitally-modulated signal) in this source.

Orbitally-modulating X-ray emission with a broad maximum at the inferior conjunction of the pulsar
(INFC; pulsar between the companion and observer) has been well detected
in bright `redbacks' (RBs; $>0.1M_{\odot}$ companion) \citep[e.g.,][]{Roberts2013,Wadiasingh2017}, and
XSS~J12270$-$4859, PSR~J2039$-$5617, and PSR~J2339$-$0533 (J1227, J2039, and J2339 hereafter)
are no exception.
These RBs are particularly intriguing because they exhibit orbital modulation in the GeV band as well
\citep[][]{ntsl+18,ark20,Clark+2021,An2022}.
Their LAT LCs have a maximum at the superior conjunction of the pulsar (SUPC).
At this phase, the sources' X-ray emission is at a minimum level, indicating that the IBS tail
is in the opposite direction of the LoS \citep[][]{kra19}.
Hence, the scenario that ascribes the gamma rays to synchrotron radiation
from the IBS electrons \citep[][]{Merwe+2020} seems implausible for these RBs.
Instead, it was suggested that inverse-Compton (IC) scattering off of the
companion's BB photons by upstream (unshocked) wind particles may generate the gamma-ray
modulation due to orbital variation of the scattering
geometry. However, in this case, energy injection from the pulsar seemed insufficient \citep[][]{ark20,Clark+2021}.
Alternatively, \citet{Merwe+2020} suggested that synchrotron radiation from the upstream particles,
which pass through the IBS and interact with the companion's $B$,
may be responsible for the gamma-ray emission \citep[see also][]{Clark+2021}.
This scenario may explain the emission strength and phasing of the gamma rays,
but it is yet unclear whether the observed LC shapes can also be reproduced.

In this paper, we explore scenarios for the gamma-ray modulations discovered in the
three RBs, J1227, J2039, and J2339, by modeling their multiband emission.
We construct X-ray spectral energy distributions (SEDs) and orbital
LCs of the targets, and supplement them with
published LAT measurements (Section~\ref{sec:sec3}).
We then apply an IBS model to the data
and investigate mechanisms for the modulating GeV emissions
from the sources in Section~\ref{sec:sec4}.
Finally, we discuss the results and present our conclusions in Section~\ref{sec:sec5}.

\section{X-ray Data Analysis}
\label{sec:sec2}
We analyze the X-ray observations of the targets (Table~\ref{ta:ta1}),
taking into account the potential BB emission from the pulsar.
This consideration is crucial because the BB emission may
significantly impact the measurements of the nonthermal IBS X-ray spectrum and LC.

\newcommand{\marka}{\tablenotemark{\tiny{\rm a}}}
\newcommand{\markb}{\tablenotemark{\tiny{\rm b}}}
\begin{table}
\begin{center}
\caption{X-ray data used in this work}
\label{ta:ta1}
\scriptsize{
\begin{tabular}{lccccc}
\hline \hline
Target					& Instrument   & Date    & Exposure & Comment \\  
					&              & (MJD)   & (ks) & \\  \hline
J1227					& NuSTAR  &  59426  & 100 & $\cdots$  \\  \hline
J2039					&  XMM    &  56575  & 43/43/37  & M1/M2/PN\marka \\  \hline
\multirow{4}{*}{J2339}			& Chandra &  55118  & 21  & $\cdots$  \\ 
					&  XMM\markb &  56640  & 41/42  & M1/M2 \\ 
					&  XMM\markb &  57745  & 86/85  & M1/M2 \\ 
					&  NuSTAR &  57744  & 154 & $\cdots$  \\ \hline
\end{tabular}}\\
\end{center}
\footnotesize{
\marka{M1: Mos1. M2: Mos2.}\\
\markb{XMM13 and XMM16 for the earlier and later observation, respectively (Figure~\ref{fig:fig1}).}\\
}
\end{table}

Although there are archival X-ray observations of J1227
taken just after the source's state transition in late 2012 \citep[$\sim$MJD~56250;][]{Bassa+14},
we do not include them in our analysis. This is due to the potential effects of a residual accretion disk,
as indicated by strong variability in the source's flux and LC shape
observed between 2013 and 2015 \citep[][]{deMartino+15}.
Consequently, we acquired new NuSTAR data in 2021, $\ge$8\,yrs after the transition.
The XMM observation of J2039 was analyzed
by \citet{Romani2015} and \citet{Salvetti2015}; however, we reanalyze the data to accurately
measure the `nonthermal' emission from the source. This reanalysis involves a careful examination
of contamination from the pulsar's BB emission. The data for J2339 were analyzed by
\citet{kra19}, with a focus on phase-resolved spectroscopy and LCs.
We reanalyze the data to construct a phase-averaged SED.

\begin{figure*}
\centering
\includegraphics[width=7. in]{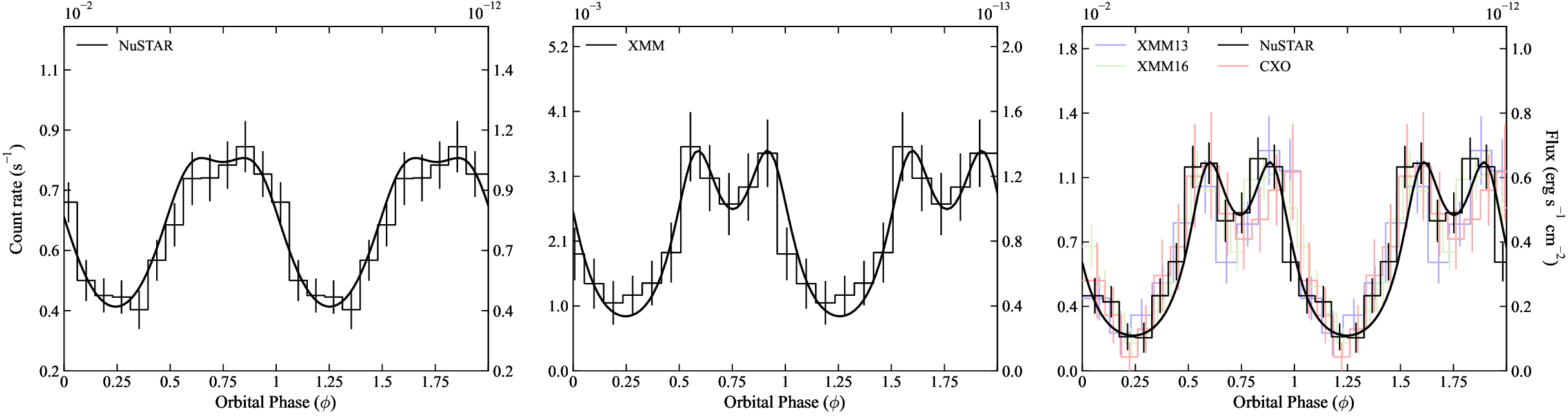} 
\figcaption{Exposure-corrected LCs of J1227 (left), J2039 (middle), and J2339 (right).
The left and right ordinates show the count rates and fluxes, respectively, where
the latter were estimated by comparing the phase-averaged fluxes to the observed counts.
We used the 2--10\,keV band for the Chandra and XMM LCs, and the 3--20\,keV band for the NuSTAR LCs.
The solid curves are LCs computed with our IBS model for Scenarios~1 and 2
(see Sections~\ref{sec:sec3_2} and \ref{sec:sec4_3}).
\label{fig:fig1}
}
\vspace{0mm}
\end{figure*}

\subsection{Data reduction}
\label{sec:sec2_1}
We processed the XMM data using the {\tt emproc} and {\tt epproc} tools integrated in SAS~2023412\_1735.
We further cleaned the data to minimize contamination by particle flares.
Note that we did not use XMM timing-mode data due to the low signal-to-noise ratio.
The Chandra observation was reprocessed with the {\tt chandra\_repro} tool of CIAO~4.12, and
the NuSTAR data were processed with the {\tt nupipeline} tool in HEASOFT~v6.31.1 using
the {\tt saa\_mode=optimized} flag.
For the analyses below, we employed circular source regions with radii of $R=3''$, $R=16''$ and $R=30''$
for Chandra, XMM, and NuSTAR, respectively.
Backgrounds were extracted from source-free regions with radii of $R=6''$ for Chandra,
$R=32''$ (or $R=16''$ for small-window data) for XMM, and $R=45''$ for NuSTAR.

\subsection{X-ray light curves}
\label{sec:sec2_3}
To generate orbital LCs of the sources for use in our modeling (Section~\ref{sec:sec4}),
we barycenter-corrected the photon arrival times and folded them using the orbital period ($P_{\rm B}$)
and the time of the ascending node ($T_{\rm ASC}$) measured for each source
\citep[][]{rrbs+15,Clark+2021,ark20}.
We should note that the exposures of the observations are not integer multiples of the orbital periods,
resulting in uneven coverage of orbital phases. Furthermore, there are observational gaps
caused by flare-removal (XMM) and Earth occultation (NuSTAR),
which also introduce nonuniformity in the phase coverage.

We investigated the effects of the nonuniform phase coverage and found that
the observational gaps present in the XMM and NuSTAR data were randomly distributed in phase,
resulting in spiky features in the LCs.
However, these random variations did not significantly impact the overall flux measurements, which
remained within a range of $\lapp$1--2\%.
In the cases of the Chandra LC of J2339 and the XMM LC of J2039, systematic trends were observed.
The Chandra observation had $2\times$ more exposure for a phase interval when
the source appeared bright, while the XMM data had $\sim$50--60\% longer exposure near the minimum phase.
Due to the significant distortion of the LCs caused by both random and systematic exposure variations,
we corrected the LCs for the unequal exposure.

We computed the source exposures using the good time intervals of the observations,
folded the exposures on the orbital periods, and divided the LCs by the folded exposure
to account for the exposure variations.
The exposure-corrected LCs are presented in Figure~\ref{fig:fig1}, where
they exhibit a single- or double-peak structure.
To minimize contamination from the orbitally-constant BB emission (Section~\ref{sec:sec2_2})
while ensuring high signal-to-noise ratios,
we utilized the 2--10\,keV and 3--20\,keV bands for the Chandra/XMM and NuSTAR LCs, respectively.
The LC of J1227, measured with the 2021 NuSTAR data, displays a broad single bump (Figure~\ref{fig:fig1} left)
similar to the 2015 NuSTAR LC \citep[][]{deMartino+20}. This similarity suggests that
the LC of J1227 has remained stable since 2015. It is worth noting that the LC of this source
exhibited significant morphological changes between 2013 and 2015,
immediately after the state transition \citep[][]{Bogdanov+14,deMartino+15}.
The multi-epoch LCs of J2339 appear almost the same \citep[Figure~\ref{fig:fig1} right; see also][]{kra19},
indicating that the source has been stable for $\sim$2600\,days.

\begin{figure*}
\centering
\includegraphics[width=7 in]{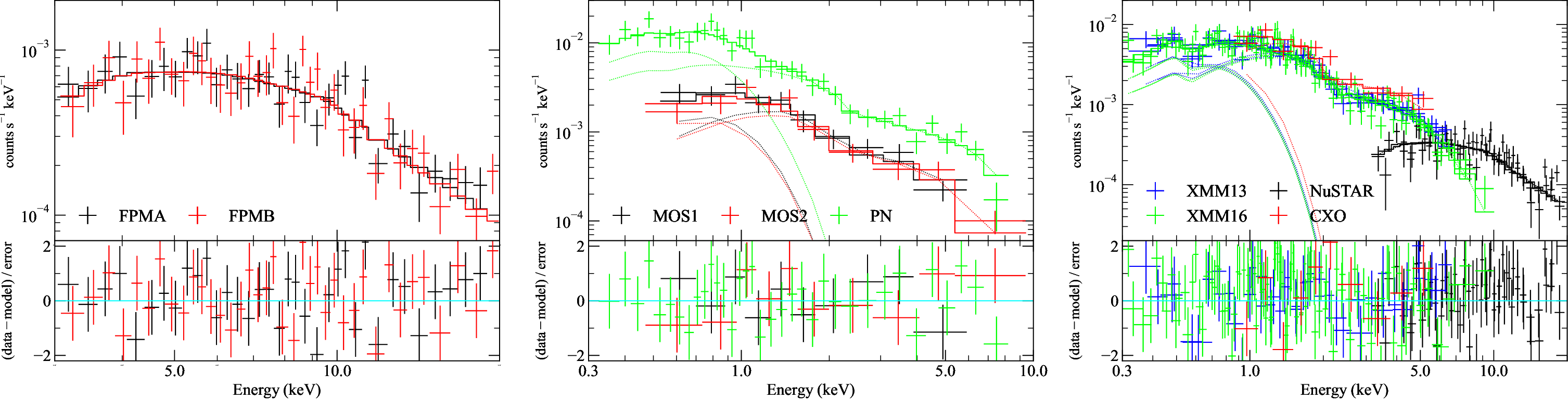}
\figcaption{X-ray spectra of J1227 (left), J2039 (middle), and J2339 (right).
The solid lines represent the best-fit {\it XSPEC} models, and the bottom panels display the
fit residuals.
\label{fig:fig2}
}
\vspace{0mm}
\end{figure*}

\subsection{Spectral analysis}
\label{sec:sec2_2}
As mentioned previously, the random variations in exposure do not pose a concern
for spectral analyses. However, the systematic excesses in exposure in the Chandra
(J2339) and the XMM (J2039) data affected the spectral measurements.
Hence, we removed the time intervals corresponding to the excess exposure from the data,
which resulted in a $\sim$10\% reduction in the livetime.

We generated the X-ray spectra of the targets and created corresponding
response files using the standard procedure suitable for each observatory.
The spectra were grouped to contain at least 20 events per bin and were then fit with
an absorbed BB or power-law (PL) model using the $\chi^2$ statistic.
For Galactic absorption, we employed the {\tt tbabs} model
with the {\tt wilm} abundances \citep[][]{wam00} and the {\tt vern} cross section \citep[][]{vfky96}.
The hydrogen column densities ($N_{\rm H}$) for these high-latitude sources were
inferred to be low, but their values were not well constrained
by the fits \citep[see also][]{deMartino+20,Romani2015,kra19}.
Therefore, we held $N_{\rm H}$ fixed at the values estimated
based on the HI4PI map \citep[][]{HI4PI2016}.\footnote{https://heasarc.gsfc.nasa.gov/cgi-bin/Tools/w3nh/w3nh.pl}

Below, we present results of our spectral analysis for each source.

{\bf J1227}: The new 2021 NuSTAR spectra of J1227 were adequately described by a simple PL
(Figure~\ref{fig:fig2} and Table~\ref{ta:ta2}), and the results are very similar
to the previous ones obtained from the 2015 NuSTAR data
\citep[see][]{deMartino+20}. However, the source flux was $\sim$20--30\% lower in 2021 compared to 2015.
Although this flux difference is not highly significant ($\lapp3\sigma$),
we use the new results for our modeling (Section~\ref{sec:sec4}).
We thoroughly examined the archival XMM data acquired in 2013--2014 (after the transition)
to identify a potential BB component, but no significant evidence for it 
was found, as previously noted by \citet{deMartino+15}.

{\bf J2039}: For this source, 
the PL model was acceptable, whereas the BB model was statistically
ruled out. Our PL results are consistent with
the previous measurements \citep[][]{Romani2015,Salvetti2015}.
However, we noticed a trend in the fit residuals.
To address this, we fit the spectrum with a BB+PL model (Figure~\ref{fig:fig2}).
We then conducted an $f$ test to compare the PL and BB+PL models,
and found that the BB+PL model was favored over the PL model with an $f$-test probability
of $5\times 10^{-6}$; we verified this result using simulations.
We report the best-fit parameters of the BB+PL model in Table~\ref{ta:ta2}.
\citet{Salvetti2015} also favored the BB+PL model, but they
did not report the best-fit parameter values.

\begin{table}
\begin{center}
\caption{Spectral fit results}
\label{ta:ta2}
\scriptsize{
\begin{tabular}{lcccccc}
\hline
Target & Energy  & $kT$     & ${R_{\rm BB}}$\marka & $\Gamma_X$ & ${F_X}$\markb  & $\chi^2$/dof \\  
       & (keV)   & (keV)    &  (km)        &     &     &  \\  \hline
J1227  &  3--20  & $\cdots$ & $\cdots$ & 1.28(7)  & 3.67(18) & 71/70  \\ 
J2039  & 0.3--10 & 0.17(2)  & 0.24(4)  & 1.07(12) & 0.75(9) & 35/49 \\ 
J2339  & 0.3--20 & 0.14(1)  & 0.31(4)  & 1.10(5)  & 1.60(11) & 236/217 \\  \hline
\end{tabular}}
\end{center}
\footnotesize{Note. Numbers in parenthesis represent the 1$\sigma$ uncertainty.\\
\marka{For assumed distances of 1.37\,kpc, 1.7\,kpc, and 1.1\,kpc for
J1227, J2039, and J2339, respectively \citep[][]{Jennings2018,Clark+2021}.}\\
\markb{Absorption-corrected PL flux in the 3--10\,keV band in units of $10^{-13}$\,\fluxcgs.}\\
}
\end{table}

{\bf J2339}:
We jointly fit the Chandra, XMM, and NuSTAR spectra of J2339,
allowing a cross normalization for each spectrum to vary.
Both the BB and PL models were statistically ruled out
with the $\chi^2$ probabilities of $<10^{-5}$,
and we observed a discernible trend in the residual, similar to the case of J2039.
Therefore, we fit the data with a BB+PL model, and this model provided an
acceptable fit with the $\chi^2$ probability of 0.18.
This finding significantly strengthens the previous suggestion of BB emission from the source
\citep[$f$-test probability of $10^{-3}$;][]{Yatsu2015}.

\begin{figure*}
\centering
\includegraphics[width=7. in]{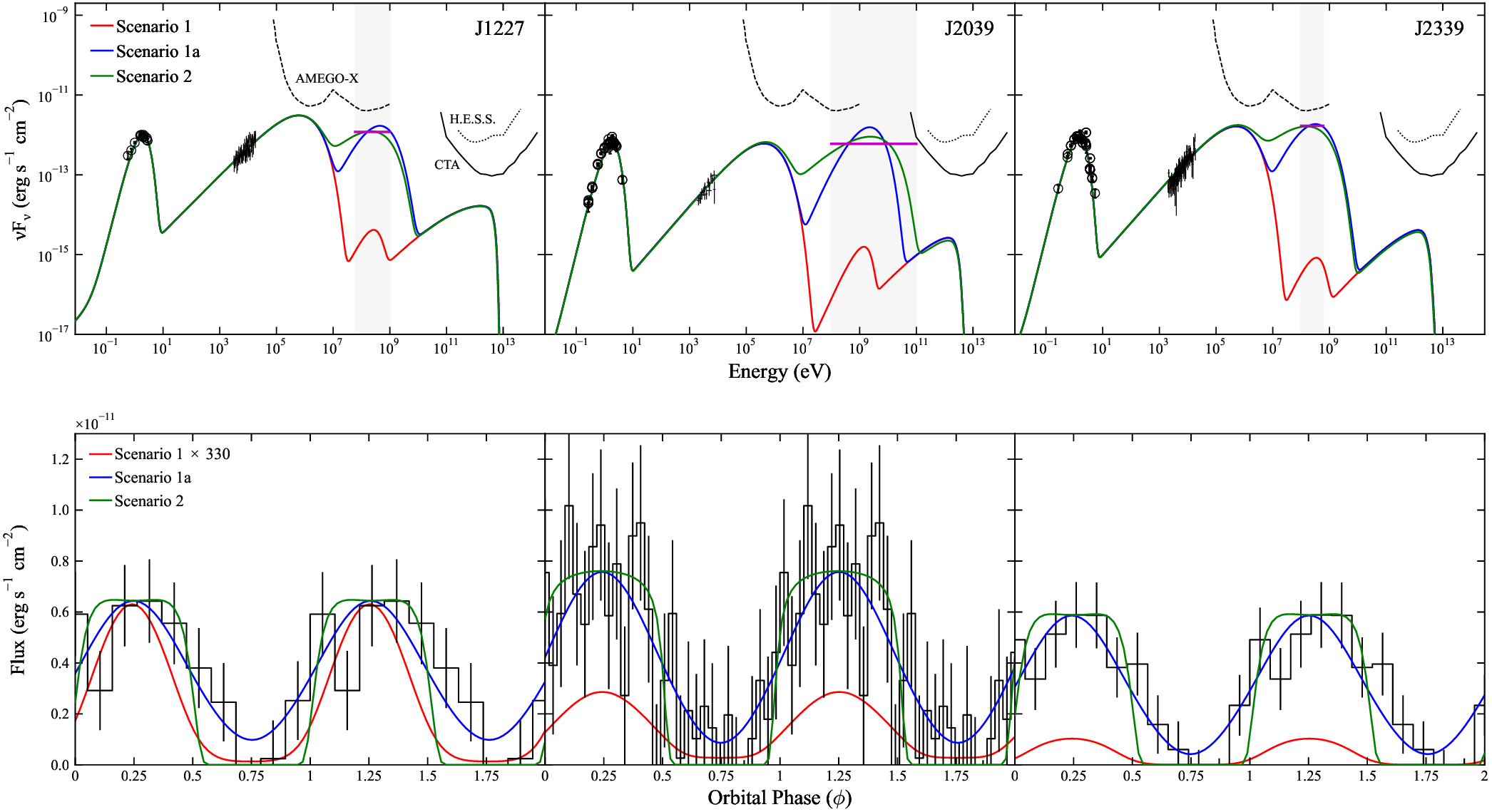}
\figcaption{Broadband SEDs (top row) and LAT LCs (bottom row)
of J1227 (left), J2039 (middle), and J2339 (right).
The X-ray data points are our measurements,
and the optical and LAT data are taken from the literature (see text).
We present models based on Scenarios~1 and 2 (Section~\ref{sec:sec3_2}) as red
and green curves, respectively. Blue curves display the case that the wind in Scenario~1
is arbitrarily decelerated (Scenario~1a).
({\it Top}): The magenta horizontal lines indicate
the flux levels for the modulating signals estimated based on the modulation fractions
of the LCs and pulsar fluxes. We also show the AMEGO-X (black dashed curve),
H.E.S.S. (black dotted curve), and CTA (south 50h; black solid curve)
sensitivities for reference.
({\it Bottom}): We subtracted constant levels from the LC data.
For legibility, we increased the amplitude of the model-computed LC for Scenario~1 by a factor of 330.
\label{fig:fig3}
}
\vspace{0mm}
\end{figure*}

\section{Multiband data and emission scenarios}
\label{sec:sec3}

\subsection{Broadband SED and LC data}
\label{sec:sec3_1}
We construct the broadband SEDs and LCs of our target RBs (Table~\ref{ta:ta1}). Our analysis
of the X-ray data provided nonthermal X-ray SEDs and LCs (Figure~\ref{fig:fig1} and
Table~\ref{ta:ta2}). For the modeling, we converted the count units of the
X-ray LCs into flux units by comparing the phase-averaged flux to the observed counts for each source.
For the LAT data, we adopted the published results \citep[][]{ntsl+18,ark20,Clark+2021,An2022}.
These previous analyses used different energy bands:
60\,MeV--1\,GeV for J1227, 100\,MeV--100\,GeV for J2039, and 100\,MeV--600\,MeV for J2339.
To ensure consistency, we converted the count units of the LCs into flux units using
the LAT models provided in the Fermi-DR3 catalog \citep[][]{fermi4fgldr3}.
Additionally, we subtracted constant levels from the gamma-ray LCs, assuming that
the constant emission originates from the pulsar's magnetosphere
rather than the IBS or upstream wind \citep[see also][]{Clark+2021}.

Since the spectrum of the orbitally-modulated LAT signals
has not been well measured, we present in Figure~\ref{fig:fig3}
flux levels (horizontal lines) estimated by scaling down the pulsar SEDs
according to the modulation fractions of the LAT LCs (typically $\sim$30\%).
The actual IBS spectra might be softer than those of the pulsars as
the observed modulations of the targets were more pronounced at low energies \citep[e.g.,][]{ntsl+18,ark20,An2022}.
Therefore, the gamma-ray SEDs of the modulating signals are likely to peak at $\le$\,GeV energies
\citep[see also Figure~4 of][]{Clark+2021}.
The spectra of the optical companions, which provide seeds for IC emission,
were obtained from the literature \citep[][]{Rivera+2018,kra19,Clark+2021} and the {\tt VizieR}
photometry database.\footnote{http://vizier.cds.unistra.fr/vizier/sed/}
These spectra represent the observed emission near the optical-maximum phase.
The broadband SEDs and gamma-ray LCs of the targets are displayed in Figure~\ref{fig:fig3}.

\begin{figure*}
\centering
\includegraphics[width=6.8 in]{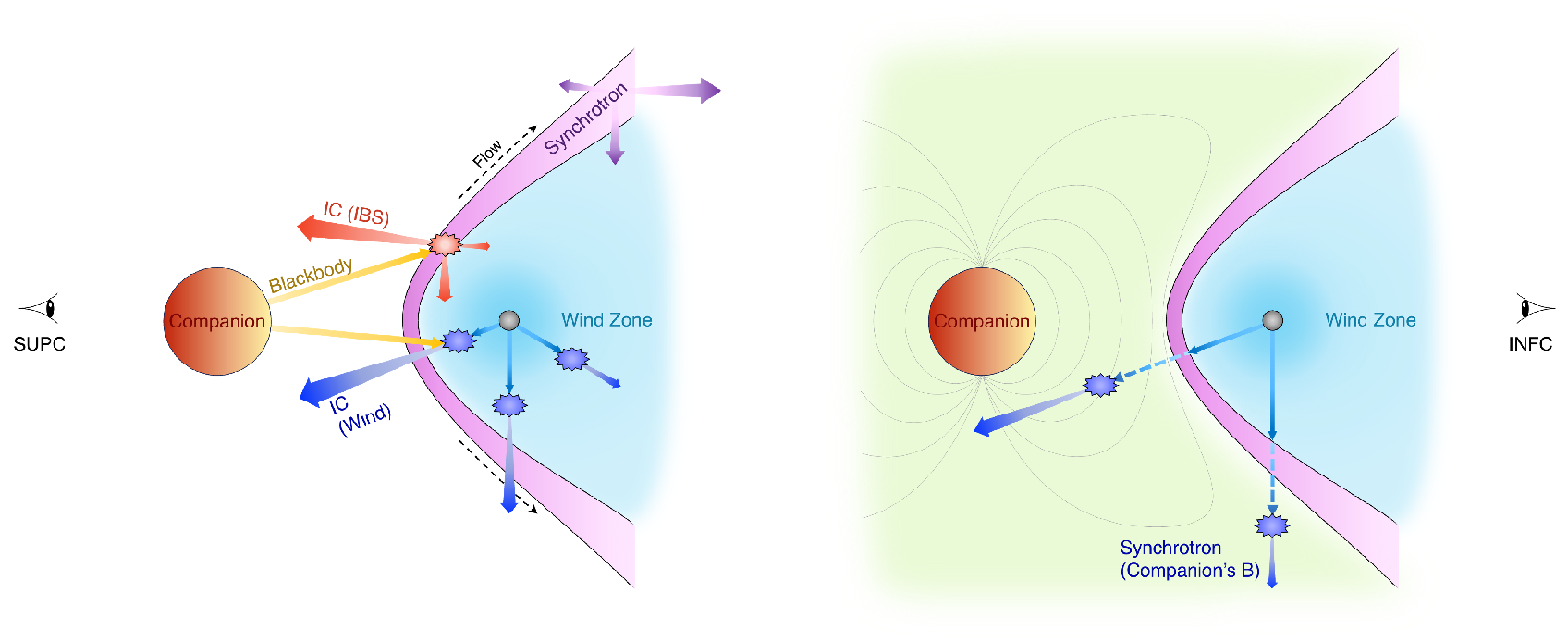}
\figcaption{Cartoons that depict emission scenarios employed in this work.
({\it Left}): Emission components common to Scenarios~1 and 2. Purple, red, and blue
arrows denote synchrotron, IC-in-IBS, and IC-in-wind emissions, respectively.
Note that the synchrotron emission is stronger in the flow direction due to relativistic aberration.
For each emission component, three representative directions (toward INFC, SUPC, and in between)
are displayed, where thicker arrows mean stronger emission.
({\it Right}): Synchrotron radiation under the companion's $B$ (green region) by the wind particles that penetrate
the IBS. For Scenario~2, we add this emission component to those of Scenario~1.
\label{fig:fig4}
}
\vspace{0mm}
\end{figure*}

\subsection{Emission scenarios}
\label{sec:sec3_2}
In IBS scenarios, a relativistic electron/positron plasma (advected in the MHD pulsar wind) originating from a pulsar
is injected into an IBS formed by wind-wind or wind-$B$ interaction.
The electrons (electrons+positrons) are accelerated at the shock and flow along the IBS.
These IBS electrons emit synchrotron radiation in the X-ray band, which is
Doppler-boosted along the bulk flow direction (Figure~\ref{fig:fig4}).
The companion provides seeds for IC scattering to the electrons in the IBS and in the
pulsar-wind region. While this scenario has been successful in modeling the X-ray SEDs and LCs
of pulsar binaries \citep[][]{rs16,Wadiasingh2017,kra19,Merwe+2020,Cortes2022},
it was suggested that the scenario cannot explain the recently-discovered GeV modulations
from our RB targets \citep[][see Section~\ref{sec:sec3_2_1}]{ark20,Clark+2021}.
We check this basic scenario (Scenario~1) to confirm
the previous suggestion, and we adjust the parameters within this scenario
to offer a phenomenological explanation for the data.
Then, we explore an alternative scenario to explain
the LAT measurements. Note that these scenarios share the same
mechanism for the X-ray emission (synchrotron radiation from IBS electrons), and
therefore, our descriptions of the scenarios concentrate on the gamma-ray emission mechanisms.

\begin{itemize}
\item {\it Scenario 1:} This is the basic IBS scenario where electrons in the
cold wind and in the IBS IC-upscatter the companion's BB photons
to produce gamma-ray emission (Figure~\ref{fig:fig4} left).
\item {\it Scenario 2:} It was suggested that a sufficiently energetic component of the upstream
pairs passes through the IBS unaffected and emits synchrotron radiation
under the influence of the companion's $B$ \citep[][]{Merwe+2020},
producing GeV gamma rays (Figure~\ref{fig:fig4} right).
This component is added to the gamma-ray flux of Scenario~1.
\end{itemize}

To summarize, we consider three emission zones as listed below.
\begin{itemize}
\item {\it Wind zone (cyan in Figure~\ref{fig:fig4}):}
This is an emission zone between the pulsar's light cylinder and the IBS.
Electrons in this zone are assumed to be cold (i.e., $\delta$ distribution) and relativistic
(but see Section~\ref{sec:sec4_1_1} for Scenario\,1a).
In our phenomenological study, we consider only the IC emission from the electrons, assuming that
their synchrotron emission is weak (but see Section~\ref{sec:sec5}).
\item {\it IBS zone (pink in Figure~\ref{fig:fig4})}: The electrons in the wind zone
are injected into this IBS zone, and thus the number and energy of the electrons in this
zone are connected to those in the wind zone (Equations~(\ref{eq:N0})--(\ref{eq:etaratio}) below).
In this IBS zone, shock-accelerated electrons flow along the IBS surface,
and they produce both synchrotron and IC emissions.
We do not consider synchrotron-self-Compton emission
from this zone, as its flux has been assessed to be
negligibly small \citep[][]{Merwe+2020}.
\item {\it Companion zone (green in Figure~\ref{fig:fig4} right):}
This emission zone is used only for Scenario~2.
Most of the upstream pairs interact with the IBS zone, but a sufficiently energetic component of them with large gyro radii are assumed to pass through the IBS unaffected and reach
this zone. These electrons can produce both synchrotron and IC emission with the former dominating over the latter due to the
strong $B$ of the companion. Therefore, we consider only the synchrotron emission.
\end{itemize}

We investigate the fundamental properties of the emission scenarios using
simplified calculations in Sections~\ref{sec:sec3_2_1}--\ref{sec:sec3_2_3},
where, for simplicity, we assume that the energy distributions of the particles
and their radiation are $\delta$ functions and that the IC scattering
occurs in the Thomson regime.
We also neglect Doppler boosting caused by the bulk motion of
the IBS flow \citep[e.g.,][]{Wadiasingh2017} and carry out our analytic investigation
in the flow rest frame (equivalent to the observer frame in this section),
as Doppler factors of IBS flows
in pulsar binaries have been inferred to be small \citep[e.g.,][see also Table~\ref{ta:ta4}]{rs16,kra19}.
These analytic calculations, made utilizing a one-zone approach and mono-energetic distributions,
provide rough estimates of the model parameters
that will be used as inputs for detailed multi-zone IBS computations performed without
the aforementioned assumptions (Section~\ref{sec:sec4}).

\subsubsection{Scenario 1: Basic IBS scenario}
\label{sec:sec3_2_1}
The energy distribution of electrons in IBSs is often assumed to be a power law
$\frac{dN}{d\gamma_e dt} \propto \gamma_e^{-p_1}$.
Since a pulsar supplies energy to the IBS, we require
\begin{equation}
\label{eq:energy}
\int_{\gamma_{s,\rm min}}^{\gamma_{s,\rm max}} \gamma_e m_e c^2 \frac{dN}{d\gamma_e dt} d\gamma_e
= \eta_s \dot E_{\rm SD} f_{\Omega},
\end{equation}
where $\eta_s$ represents the energy conversion efficiency of the IBS,
$\dot E_{\rm SD}$ is the pulsar's spin-down power,
and $f_{\Omega}$ (assumed to be 1 in this section) is the fraction of the solid angle subtended by the IBS.
A fraction ($\eta_\gamma$) of $\dot E_{\rm SD}$ is converted to the pulsar's gamma-ray radiation
\citep[Table~\ref{ta:ta3}; see also][]{fermi2PC}, while the remaining energy
is eventually converted into the particles' energy (fraction $\eta_s$) and the magnetic energy
(fraction $\eta_B$). We assume that the magnetic energy and radiative energy loss are negligibly small within the
IBS \citep[e.g.,][]{kc84a}.

For a mono-energetic electron distribution
$\frac{dN}{d\gamma_e dt} = \dot N_s \delta(\gamma_e - \gamma_s)$,
Equation~(\ref{eq:energy}) can be rewritten as
\begin{equation}
\label{eq:monoenergy}
\dot N_s \gamma_s m_e c^2 = \frac{N_s}{t_s} \gamma_s m_e c^2 = \eta_s \dot E_{\rm SD},
\end{equation}
where $t_s$ is the residence time in the emission zone. It is given by
$l_s/v_{\rm IBS}$, where $l_s$ is the length of the IBS (Figure~\ref{fig:fig4})
and $v_{\rm IBS}$ represents the bulk-flow speed within the IBS.
The synchrotron-emission frequency of these electrons is given by
\begin{equation}
\label{eq:synu}
h\nu_{\rm SY}\approx 1.6\times 10^{-11} \left (\frac{B}{1\rm\ G} \right ) \gamma_s^2\rm \ keV,
\end{equation}
and the observed flux would be
\begin{equation}
\label{eq:Fsy}
F_{\rm SY} = \frac{c \sigma_{\rm T}N_s}{3\pi d^2} \gamma_s^2 U_B,
\end{equation}
where $\sigma_{\rm T}$ is the Thomson scattering cross section, $U_B\equiv \frac{B^2}{8\pi}$
is the magnetic energy density, and $d$ is the distance between the observer and the source.
For computation of the flux in a certain energy band, particle cooling needs to be considered.
$t_s$ is typically shorter than the cooling timescale in IBSs of pulsar binaries
\citep[Table~\ref{ta:ta4}; see also][]{Merwe+2020}.
So we assume the emission timescale in the IBS ($\tau_s$) to be approximately the residence time.
By combining Equations~(\ref{eq:monoenergy})--(\ref{eq:Fsy}), we obtain
\begin{equation}
\label{eq:B}
B \approx 0.6 \left ( \frac{d_{\rm kpc}^2 F_{\rm X,-13}}{\eta_s \dot E_{\rm SD, 35} \tau_s}\right )^{2/3}
\left ( \frac{h\nu_{\rm SY}}{1\rm \ keV} \right )^{-1/3}\rm \ G
\end{equation}
in the IBS, where $d_{\rm kpc}$ is $d$ in units of kpc, $F_{\rm X,-13}$ is
the X-ray flux in units of $10^{-13}$\,\fluxcgs, and $\dot E_{\rm SD, 35}$ is
$\dot E_{\rm SD}$ in units of $10^{35}\rm \ erg \ s^{-1}$.

In IBS scenarios, gamma-ray emission is assumed to arise from IC processes
involving electrons in the IBS and wind zones (Figure~\ref{fig:fig4}).
The emission power generated by IC scattering between an electron with
a Lorentz factor $\gamma_e$ ($\gg 1$) and seed photons with energy density $u_*$ and frequency $\nu_*$
is given by
\begin{equation}
\label{eq:IC}
P_{\rm IC} = \sigma_{\rm T} c u_* (1-\beta_e \mu)^2\gamma_e^2,
\end{equation}
\citep[e.g.,][]{d13}, where $\beta_e = \sqrt{1 - 1/\gamma_e^2}\approx 1$
and $\mu$ is the cosine of the scattering angle ($\mu=-1$ for head-on collisions).
$\mu$ is one of the key parameters that determines
the shape and phasing of the gamma-ray LCs.
The observed frequency of the IC-upscattered photons is given by
\begin{equation}
\label{eq:ICnu}
h\nu_{\rm IC} = \gamma_e^2(1 - \beta_e \mu)^2 h\nu_*.
\end{equation}
The flux of the IC emission from the IBS electrons can be determined by 
\begin{equation}
\label{eq:Fic0}
F_{\rm IC,IBS} = F_X \frac{u_*}{U_B},
\end{equation}
independent of the number of electrons ($N_s$) in the IBS, as the same electrons
produce both synchrotron and IC emission.

On the contrary, to estimate the IC flux from the cold `wind' particles
(see Equation~(\ref{eq:Fic}) below), we need to determine their number.
Assuming a $\delta$ distribution for the cold wind particles, we have
\begin{equation}
\label{eq:deltaenergy}
\frac{dN}{d\gamma_e dt}=\dot N_w\delta(\gamma_e-\gamma_w)
\end{equation}
and
\begin{equation}
\label{eq:Efrac}
\gamma_w \dot N_w m_e c^2 = \eta_w \dot E_{\rm SD},
\end{equation}
where $\gamma_w$ represents the Lorentz factor of the upstream particles, $\dot N_w$ is
the number of particles injected (per second) by the pulsar into the wind zone (cyan region in Figure~\ref{fig:fig4}),
and $\eta_w$ ($<1$) is an efficiency factor
that accounts for the conversion of the pulsar's energy output into
particles within the wind zone.
The total number of particles in the wind zone is then given by
\begin{equation}
\label{eq:Nw}
N_w = \dot N_w t_w = \dot N_w \frac{l_w}{v_{\rm wind}},
\end{equation}
where $t_w$ is the residence time, $l_w$ is the size of the zone (cyan in Figure~\ref{fig:fig4}),
and $v_{\rm wind}$ is the bulk speed of the upstream wind
with $v_{\rm wind}=c\sqrt{1-1/\gamma_w^2}$ in this `cold-wind' case.
Because the upstream wind pairs are subsequently injected into the IBS and the $B$
energy is further converted to particle energy in the IBS \citep[][]{kc84a, Sironi2011},
we require
\begin{equation}
\label{eq:N0}
\dot N_s = \dot N_w,
\end{equation}
and
\begin{equation}
\label{eq:etasw}
\eta_s \ge \eta_w.
\end{equation}
as magnetic energy may not be fully dissipated in the wind zone, and radiative energy losses in the IBS are negligible.
These equations imply
\begin{equation}
\label{eq:etaratio}
\frac{\eta_w}{\eta_s} = \frac{\gamma_w}{\gamma_s} \le 1.
\end{equation}
We should note that Equation~(\ref{eq:etaratio}) is applicable exclusively to
mono-energetic distributions for a representative spatial zone.
For arbitrary phase space distributions, one should substitute $\gamma_w$ and $\gamma_s$ with their spatial and momenta averages in the volume of interest (Section~\ref{sec:sec4}).
For homogeneous one-zone models, as considered here, Equation~(\ref{eq:etaratio}) involves calculating the averages using the expression: $\int \gamma \frac{dN}{d\gamma dt} d\gamma/\int\frac{dN}{d\gamma dt}d\gamma$.

By combining Equations~(\ref{eq:IC}), (\ref{eq:ICnu}), (\ref{eq:Efrac}) and (\ref{eq:Nw}), the IC flux
of the upstream particles, e.g., in the case of head-on collisions, can be computed as
\begin{equation}
\label{eq:Fic}
\begin{aligned}
& F_{\rm IC,wind} =  4\sigma_{\rm T} c u_* \gamma_w^2 \frac{\eta_w \dot E_{\rm SD} \tau_w}{4\pi d^2 \gamma_w m_e c^2} \approx & \\
& 10^{-16} \frac{\eta_w \dot E_{\rm SD,35} \gamma_w}{d_{\rm kpc}^2} \left ( \frac{u_*}{1\rm \ erg\ cm^{-3}} \right )
\left ( \frac{\tau_w}{1\rm \ s} \right ) \rm \ erg\ s^{-1}\ cm^{-2}, &
\end{aligned}
\end{equation}
where $\tau_w$ is the emission timescale in the wind zone.
In the case of the cold wind with $v_{\rm wind}\approx c$ in this scenario (Scenario 1),
the residence time is $t_w\approx$1\,s, which is shorter than
the cooling timescale of electrons with $\gamma_w\lapp 10^8$.
Since $\gamma_w$ is expected to be $\approx 10^4$ in this scenario (see below),
we assume $\tau_w\approx t_w$.

\begin{table}
\begin{center}
\caption{Basic parameters for J1227, J2039, and J2339}
\label{ta:ta3}
\scriptsize{
\begin{tabular}{ccccc}
\hline
Property	&	Unit	&	J1227	&	J2039	&	J2339  \\  \hline
$\dot E_{\rm SD}$	&$10^{34}\rm \ erg\ s^{-1}$	&	9.0	&	2.5	&	2.3  \\
$\eta_\gamma$	&	&	0.05	&	0.21	&	0.18  \\
$P_{\rm B}$	&	day	&	0.288	&	0.228	&	0.193  \\
$T_{\rm ASC}$	&	MJD	&57139.0716	&56884.9670	&55791.9182      \\
$M_*$		&$M_\odot$	&	0.27	&	0.18	&	0.32  \\
$R_*$		&$R_\odot$	&	0.29	&	0.30	&	0.35  \\
$T_*$		&	K	&	5700	&	5500	&	4500  \\
$a_{\rm orb}$	&$10^{11}\rm \ cm$&	1.5	&	1.2	&	1.2  \\
$i$			&	deg.	&	54.5	&	69	&	70  \\
$d$		&	kpc	&	1.37	&	1.7	&	1.1  \\ \hline
\end{tabular}}
\end{center}
\end{table}

These computations can be compared with the observed X-ray and LAT data
(Figure~\ref{fig:fig3}).
For an IBS that extends to the size of the orbit ($a_{\rm orb}\approx l_s$),
we find
$\tau_s\approx \frac{a_{\rm orb}}{v_{\rm IBS}}>\frac{a_{\rm orb}}{c}\approx 10$\,s.
From this, we can infer $B\approx 1$\,G (Equation~(\ref{eq:B})) and
$\gamma_s\approx 10^6$ (Equation~(\ref{eq:synu}))
from the observed X-ray SEDs with $F_{\rm SY}\gapp 10^{-13}$\,\fluxcgs\
at $h\nu_{\rm SY}\approx 10$\,keV (e.g., Figure~\ref{fig:fig3}).
The optical seeds provided by the companion have $h\nu_*\approx 1$\,eV and $u_*\approx 0.1\rm \ erg\ cm^{-3}$
at the position of the IBS ($\approx a_{\rm orb}$). Then, the IC flux of the IBS particles
would be $F_{\rm IC, IBS}\approx 10^{-13}$\,\fluxcgs\ (Equation~\ref{eq:Fic0}), which
may explain (part of) the observed LAT fluxes of the targets (Figure~\ref{fig:fig3}).
However, the peak of this IC-in-IBS emission
is expected to be in the $\sim$TeV band \citep[Equation~(\ref{eq:ICnu})
and Figure~\ref{fig:fig3}; see also][]{Merwe+2020}.
Therefore, IC-in-IBS emission cannot explain the LAT measurements.

In this scenario, additional gamma-ray emission arises from
IC scattering by upstream particles.
We can adjust $\gamma_w$ of the `cold and relativistic' upstream electrons such that their IC emission
peaks at $\lapp$GeV, which requires $\gamma_w\approx 10^4$ (Equation~(\ref{eq:ICnu}))
and $\eta_w\approx 0.01\eta_s\approx 0.01$ (Equation~(\ref{eq:etaratio})). This requires
that the wind zone is Poynting-flux dominated \citep[e.g.,][]{Cortes2022}.
The wind zone, extending from the pulsar's light cylinder to the IBS
(cyan in Figure~\ref{fig:fig4}),
is $<a_{\rm orb}$, and thus $\tau_w< \frac{a_{\rm orb}}{c}<10$\,s.
Consequently, $F_{\rm IC, wind}$ is typically $< 10^{-14}$\,\fluxcgs\ (Equation~(\ref{eq:Fic})), which
is orders of magnitude lower than the observed GeV fluxes of the targets (Figure~\ref{fig:fig3}).

This issue can be alleviated if $\tau_w$ ($\approx t_w$) becomes longer, e.g.,
by deceleration of the bulk speed of the upstream flow
(e.g., $t_w\approx \frac{a_{\rm orb}}{c}$ {\it vs} $t_w\approx \frac{l_w}{v_{\rm wind}}$ with $v_{\rm wind}\ll c$).
This means that the upstream wind is not cold any more and is thermalized (see Section~\ref{sec:sec4_1_1}).
For electrons with $\gamma_w\sim 10^4$, the cooling timescale
($>10^{4}$\,s) is longer than the residence time
if $v_{\rm wind} \ge 10$\,\kms. Therefore, we assume $\tau_w\approx t_w =l_w/v_{\rm wind}$ and
proceed to estimate the required $v_{\rm wind}$
to match the GeV flux.
Then Equations~(\ref{eq:monoenergy}), (\ref{eq:Fsy}) and (\ref{eq:Fic}) give
\begin{equation}
\label{eq:fluxratio}
\frac{F_{\rm IC}}{F_{\rm SY}}\approx 3\left ( \frac{v_{\rm IBS} l_w u_*}{v_{\rm wind}  l_sU_B} \right ) \left ( \frac{\eta_w}{\eta_s} \right )^2.
\end{equation}
Assuming $v_{\rm IBS}\sim c$, $l_s \sim l_w\sim a_{\rm orb}$, $u_*\sim 0.1\rm \ erg\ cm^{-3}$, $B\sim$0.1\,G, and
$\gamma_w/\gamma_s \sim 0.01$,
we estimate $v_{\rm wind}$ to be $\sim 10^{-3} c\approx 300$\,\kms\ if $F_{\rm IC}/F_{\rm SY}\approx 1$
(Figure~\ref{fig:fig3}).
We can accommodate various values of $F_{\rm IC}$ in this scenario by adjusting $v_{\rm wind}$ and/or $l_w$ (n.b. this is also equivalent to artificially reducing the spatial diffusion coefficient to an unphysically small value).
Furthermore, this scenario can explain the phasing of the LAT LC since
the model-predicted gamma-ray flux would be maximum at SUPC because of the favorable
scattering geometry (Figure~\ref{fig:fig4}).
However, our order-of-magnitude estimate above hinges on a considerably lower $v_{\rm wind}$ than upstream precursor regions presented in PIC simulations of pulsar wind shocks \citep[e.g.,][]{Sironi2011}.
Furthermore, the formation of strong shocks and particle acceleration (as demanded by the X-ray phenomenology) appears implausible if the speed of the pulsar wind is indeed decelerated to such a low value.
Notwithstanding these discrepancies,
we conduct more detailed computations based on this scenario (termed `Scenario~1a' in Figure~\ref{fig:fig3} and below)
in Section~\ref{sec:sec4}, where we adjust $v_{\rm wind}$ arbitrarily for a comprehensive analysis.

\begin{figure}
\centering
\includegraphics[width=3.2 in]{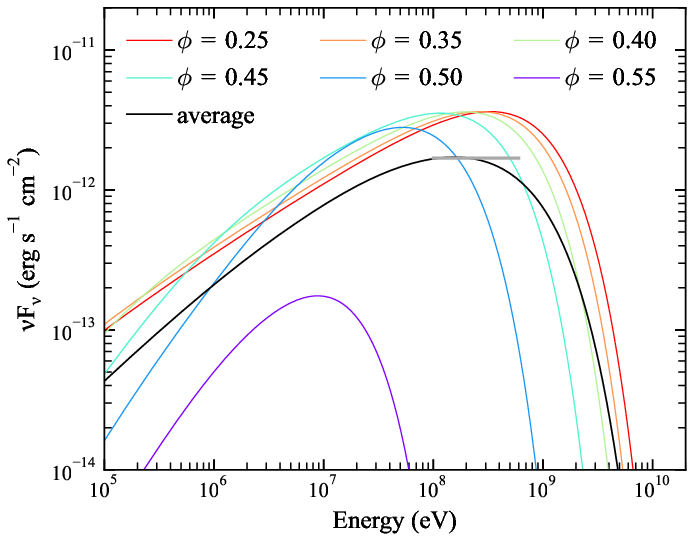}
\figcaption{Gamma-ray SEDs of J2339 at several phases computed by our numerical
model (Section~\ref{sec:sec4}) based on Scenario~2 (Section~\ref{sec:sec4}).
\label{fig:fig5}
}
\vspace{0mm}
\end{figure}

\subsubsection{Scenario 2: synchrotron radiation under the companion's $B$}
\label{sec:sec3_2_3}
Another way to increase the gamma-ray flux of RBs is to
utilize the efficient synchrotron process, as was suggested by \citet{Merwe+2020};
some of the upstream electrons may pass through the IBS
and emit gamma rays under the strong $B$ of the companion (Figure~\ref{fig:fig4} right).
If the companion has a surface $B$ of approximately kG \citep[][]{Sanchez2017,Wadiasingh2018},
electrons with a Lorentz factor $\gamma_e\approx 10^7$ can produce synchrotron
photons at $\sim$GeV energies (Equation~(\ref{eq:synu})).

Electrons with such a high Lorentz factor (primaries) can be
accelerated by the total potential drop available to the pulsar \citep[e.g.,][]{Merwe+2020} reaching $\sim 0.1$\,PeV energies.
Such high Lorentz factors proximate to the pulsar light cylinder are also required
for GeV emission in the primary curvature radiation scenario
\citep[e.g.,][]{Harding2015,Kalapotharakos2019,Harding2021,Kalapotharakos2023}.
A large fraction of the primary pairs spawns numerous secondaries with low energies, and
the remaining primaries can easily pass through the IBS unaffected since their gyro radius
is very large ($\gapp a_{\rm orb}$).
Thus, in this scenario, there are two populations of electrons in the upstream wind zone:
one $\sim 0.1$\,PeV component with a high Lorentz factor (primaries, $\gamma_p \approx 10^8$), and
the other with a lower Lorentz factor
(secondaries with $\gamma_w\approx 10^4$; referred to as `wind' in Scenario~1).
The relationships between these two populations are given as follows.
A fraction ($\eta_p$) of the pulsar's spin-down power is converted to the energy
of the primary electrons:
\begin{equation}
\label{eq:Et}
\gamma_p \dot N_p m_e c^2 = \eta_p \dot E_{\rm SD},
\end{equation}
where $\dot N_p$ represents the number of primary electrons injected (per second) by the pulsar.
Assuming a fraction ($\zeta$) of these electrons penetrates the shock,
the number of secondary electrons in the wind zone can be calculated using
\begin{equation}
\label{eq:Nt}
\dot N_w = \mathcal{M}(1 - \zeta)\dot N_p,
\end{equation}
where $\mathcal{M}$ stands for the pair multiplicity. Based on energy conservation
\begin{equation}
\label{eq:Etcons}
\eta_w \dot E_{\rm SD} = \gamma_w \dot N_w m_e c^2 = (1 - \zeta) \gamma_p \dot N_p m_e c^2,
\end{equation}
it follows that:
\begin{equation}
\label{eq:Multi}
\mathcal{M} = \gamma_p/\gamma_w
\end{equation}
and
\begin{equation}
\label{eq:etatw}
\eta_w = (1 - \zeta)\eta_p.
\end{equation}
As these secondary electrons are subsequently injected into the IBS,
Equations~(\ref{eq:N0})--(\ref{eq:etaratio}) still hold.

\begin{figure*}
\centering
\includegraphics[width=7 in]{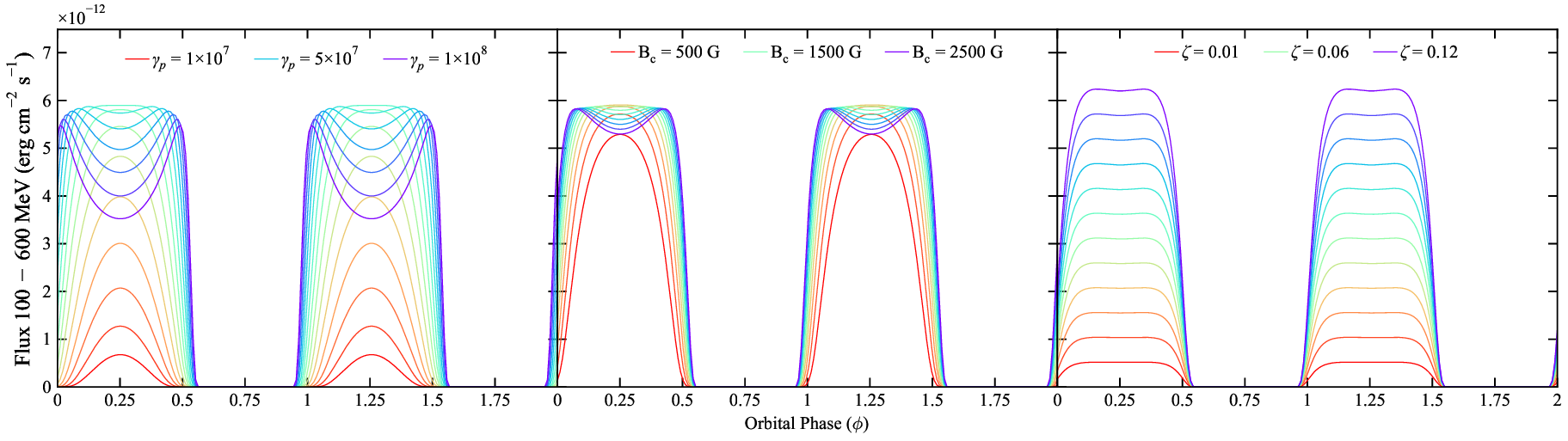}
\figcaption{GeV LCs generated through our numerical model (Section~\ref{sec:sec4}) based on Scenario~2,
employing optimized parameters specific to J2339 (Table~\ref{ta:ta4}). Changes in the LCs are
attributed to different values of $\gamma_p$ (left), $B_c$ (middle), and $\zeta$ (right), reflecting
the diverse parameter space under consideration.
\label{fig:fig6}
}
\vspace{0mm}
\end{figure*}

In this scenario, the upstream electrons are assumed to be cold. These high-energy electrons can pass through the IBS
and emit synchrotron radiation in the companion's magnetosphere.
In this case, the observed GeV flux is primarily contributed by electrons traveling along the LoS.
These electrons can interact with a strong $B$ (e.g., $\sim$kG) when in close proximity to the companion,
particularly during the SUPC phase.
The combination of high $B$ and $\gamma_p$ results in a very short synchrotron cooling time
($\ll$1\,s), given by
\begin{equation}
\label{eq:tcool}
t_{\rm cool} \approx 8 \times 10^{-4} \left (\frac{\gamma_p}{10^8}\right )^{-1} \left ( \frac{B}{0.1\rm\ kG} \right )^{-2}\rm \ s.
\end{equation}
This cooling time is much shorter than the residence time ($t_{\rm comp}=l_{\rm comp}/c$)
for any reasonable emission-zone size $l_{\rm comp}$ (e.g., $\sim a_{\rm orb}$).
So the emission timescale $\tau_{\rm comp}$ can be approximated to be $t_{\rm cool}$ during orbital phases around SUPC. Then,
the synchrotron flux arising from the companion's magnetosphere can be estimated (e.g., Equation~(\ref{eq:Fsy})) to be
\begin{equation}
\label{eq:sycomp}
\begin{aligned}
& F_{\rm SY} = \frac{c \sigma_{\rm T}  \zeta \dot N_p t_{\rm cool}}{3\pi d^2} \gamma_p^2 U_B & \\
&  \approx 8\times 10^{-10} \frac{ \zeta \eta_p \dot E_{\rm SD,35}}{d_{\rm kpc}^2} \rm \ erg\ s^{-1}\ cm^{-2}. &
\end{aligned}
\end{equation}
This scenario can explain the LAT fluxes of our targets
if $\zeta \eta_p \gapp 0.1$. Notice that $B$ does not appear in Equation~(\ref{eq:sycomp}).
This omission is a result of utilizing $t_{\rm cool}$ ($\propto B^{-2}$) for the emission timescale,
under the assumption that it is significantly shorter than $t_{\rm comp}$. This assumption is valid specifically
during orbital phases near SUPC. However, at other phases, it is more appropriate to employ $t_{\rm comp}$
instead of $t_{\rm cool}$, and the usual $B^2$ dependence of the flux is reinstated (see below).

While it might seem that this scenario does not predict orbital modulation of the gamma-ray flux
(Equation~(\ref{eq:sycomp})), changes in $B$ depending on the distance $r_*$
($B \propto r_*^{-3}$ for dipole $B$ of the companion)
between the companion and the emission zone
can induce gamma-ray modulation for two reasons.
First, the frequency of the synchrotron emission varies proportionally to $B$ for a given $\gamma_p$.
The observed flux will be high if this peak frequency falls within the
observational band, e.g., during the SUPC phase (Figure~\ref{fig:fig5}).
Second, low $B$ at certain orbital phases (e.g., large $r_*$; Figure~\ref{fig:fig4} right) increases $t_{\rm cool}$,
potentially making it longer than $t_{\rm comp}$ when $B$ is sufficiently low. In such cases,
the kinetic energy of the electrons does not fully convert into radiation
within $t_{\rm comp}$, leading to a decrease in the
emission flux (e.g., $\phi=0.55$ in Figure~\ref{fig:fig5}).
These two processes can result in a variety of LC shapes (Figure~\ref{fig:fig6} and Section~\ref{sec:sec4_1_3}).

\section{Modeling of the multiband data}
\label{sec:sec4}
The analytic exploration, performed with a one-zone approach and mono-energetic distributions,
in the previous section provides general properties of the emissions
from the RBs and establishes initial values. In this section, we leverage these findings
to conduct more detailed and precise investigations of the emissions through
our numerical model, utilizing a multi-zone approach and non-mono-energetic distributions.

\subsection{The computational methods}
\label{sec:sec4_1}
In this section, we describe the emission zones and computational methods
used in our numerical model. See \citet{KimAn2022} for a comprehensive understanding of the model components,
parameters, and their covariance.

\subsubsection{Pulsar wind zone}
\label{sec:sec4_1_1}
\begin{figure}
\centering
\includegraphics[width=3.2 in]{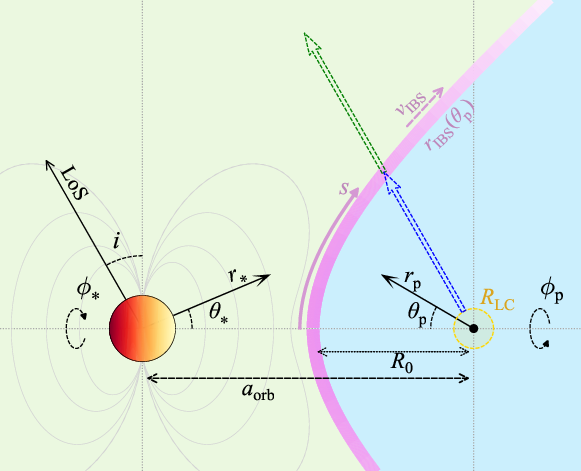}
\figcaption{Illustration showing emission zones in the vertical cross section of a system at the SUPC phase.
The companion star and pulsar are symbolized by large red and black circles, respectively (not to scale).
The thick pink curve displays the IBS.
The wind zone, situated between the pulsar's light cylinder
and the IBS (depicted in blue), contrasts with the green companion zone.
The blue and green dashed arrows represent the regions within
the wind and companion zones, respectively, where
electrons travel parallel to the LoS. Only these electrons contribute to the
observable emission due to strong Lorentz boosting.
The pulsar-centric coordinates ($r_p$, $\theta_p$, and $\phi_p$) describe emission regions,
while companion-centric coordinates ($r_*$, $\theta_*$, and $\phi_*$) detail
seed photon density, direction and the companion's $B$.
$R_0$ denotes the distance between the pulsar and the shock nose, and $s$ is the
distance along the IBS from the shock nose.
\label{fig:fig7}
}
\vspace{0mm}
\end{figure}

We assume that the pulsar wind zone (blue region in Figure~\ref{fig:fig7}) starts from the light cylinder
at a distance of $r_p=R_{\rm LC}=\frac{cP}{2\pi}$ from the pulsar, where
$P$ denotes the spin period of the pulsar.
This zone extends to the IBS surface at $r_p=r_{\rm IBS}$ marked by the pink region in Figure~\ref{fig:fig7}.

We assume that relativistic mono-energetic electrons are injected into this zone
by the pulsar (Equations~(\ref{eq:deltaenergy}) and (\ref{eq:Efrac})).
As these electrons move ballistically in the radial direction within this zone at
highly relativistic speeds, their emission is strongly beamed along
the direction of motion. Therefore, we compute IC emission
only from the electrons propagating along the LoS, denoted by blue dashed arrow in Figure~\ref{fig:fig7}.
Given that both the density of the seed photons (from the companion's BB)
and the IC scattering geometry vary based on
the location and flow direction of the emitting electrons with respect to the companion,
we divide the emission zone (essentially a line) into 100 segments.

In each segment characterized by a length of $dr_p$, we assign 
\begin{equation}
\label{eq:npreseg}
\frac{dr_p}{c}\dot N_w
\end{equation}
electrons
with the Lorentz factor of $\gamma_w$ and compute their IC emission (Section~\ref{sec:sec4_2_2}).
Note that the factor for the solid-angle fraction subtended
by the observer ($1/4\pi d^2$), which is necessary because of the approximation that the scattered photons are in the same direction as the scattering electrons, is not included in the above equation because
it is accounted for in the emission formula (Equation~(\ref{eq:ecsed})).
As previously mentioned in Section~\ref{sec:sec3_1}, the cooling timescale within this zone
exceeds the flow timescale, and thus we opt to neglect the IC cooling of the electrons.

\begin{figure}
\centering
\includegraphics[width=3.2 in]{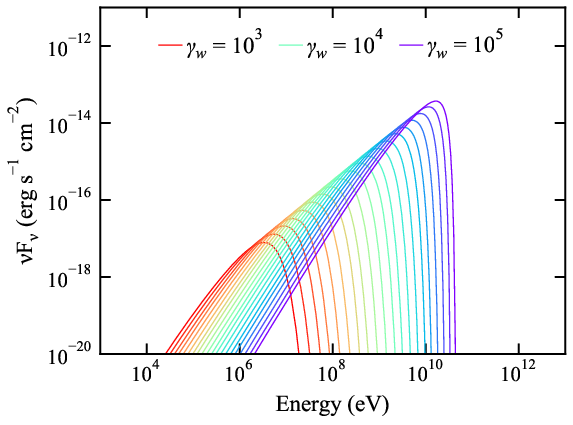}
\figcaption{IC SEDs resulting from electrons interacting with the companion's BB radiation (Equation~(\ref{eq:ecsed})).
These emissions originate specifically from the wind zone during the SUPC phase.
The displayed SEDs represent a range of values for $\gamma_w$.
All other parameters are held fixed at their optimized values for J2339 (Table~\ref{ta:ta4}).
\label{fig:fig8}
}
\vspace{0mm}
\end{figure}

The emission frequency and flux are determined by $\gamma_w$, $\eta_w$,
and the size of the wind zone, given the parameters of the pulsar, companion,
and orbit (Table~\ref{ta:ta3}).
The IC scattering geometry, the angle between the electron's motion (blue arrow in Figure~\ref{fig:fig7})
and the incident seed photons from the companion, drives orbital modulation in the GeV LC.
This effect is incorporated into the IC formulas (Section~\ref{sec:sec4_2_2}).
The size of the wind zone changes based on the IBS parameters (see below),
but this variation has little influence on the flux. Additionally, in this study, we adopt the maximum
possible value for $\eta_w$ to explore the limiting case and reduce the number of adjustable parameters.
It is worth noting that smaller values of $\eta_w$ \citep[magnetically-dominated wind; see][]{Cortes2022} can also
explain the data (Section~\ref{sec:sec4_3}). We optimize $\gamma_w$ to match the GeV flux (Figure~\ref{fig:fig8}).

Given that this basic scenario (Scenario~1) fails to yield sufficient GeV flux (Figure~\ref{fig:fig3}),
we modify this scenario by arbitrarily reducing the flow speed ($v_{\rm wind}$) in the wind zone
(increasing their residence time by decreasing their effective spatial diffusion coefficient).
In this decelerated wind case (Scenario~1a; Section~\ref{sec:sec3_2_1}), we assume a departure
from the relativistic cold wind scenario described above (Scenario~1), i.e., decelerated and thermalized, impacting
both the electron distribution, and consequently their emission.
Alongside bulk deceleration, we posit that electrons become isotropized in space and follow a
relativistic Maxwellian distribution (instead of a $\delta$ distribution) in the electron rest frame:
\begin{equation}
\label{eq:Maxwell}
\frac{dN}{d\gamma_e dt}=\dot N_w \frac{\gamma_e^2 \beta_e}{\Theta K_2(1/\Theta)} e^{-\gamma_e/\Theta},
\end{equation}
where $\beta_e=\sqrt{1-1/\gamma_e^2}$, $\Theta$ shapes the distribution,
and $K_2$ is the modified Bessel function of the second kind.
We adjust $\Theta$ to ensure that the distribution peaks at $\gamma_w$ (we use $\gamma_w$ instead of $\Theta$
as our model parameter).
Due to the isotropization, electrons traveling along the LoS exist at every point within the wind zone.
To account for this, we integrate over the entire wind zone, dividing it into 10$\times$80$\times$180
radial, polar, and azimuthal ($r_p$, $\theta_p$, and $\phi_p$ in Figures~\ref{fig:fig7}) regions.
The decelerated bulk speed in the wind zone is modeled as a constant function:
\begin{equation}
\label{eq:constv}
\vec v_{\rm wind}(r)=v_{\rm wind} \hat r_p.
\end{equation}

The number of electrons in each of the 10$\times$80$\times$180 regions is given by
\begin{equation}
\label{eq:npreslowseg}
\dot N_w \frac{dr_p}{v_{\rm wind}} f_{\Delta\Omega},
\end{equation}
where $f_{\Delta\Omega}$ represents the solid-angle fraction of the emission region.
We calculate the IC emission from these electrons,
accounting for Doppler beaming and anisotropic IC scattering (Section~\ref{sec:sec4_2}).
We should note that the energy conservation for these ``non-mono-energetic'' electrons
\begin{equation}
\label{eq:energint}
\int \gamma_e m_e c^2 \frac{dN}{d\gamma_e dt} = \eta_w \dot E_{\rm SD}
\end{equation}
requires (c.f., Equation~(\ref{eq:Efrac})) modification
by a factor of $1/\sqrt{1-(v_{\rm wind}/c)^2}$ due to the bulk motion. However, we neglect
this effect as it is less than 1\% for the estimated $v_{\rm wind}\approx 1000$\,\kms.
Additionally, we ignore the IC cooling of the electrons, as its impact is insignificant (Section~\ref{sec:sec3_1}).

\subsubsection{IBS zone}
\label{sec:sec4_1_2}
\begin{figure}
\centering
\includegraphics[width=3.2 in]{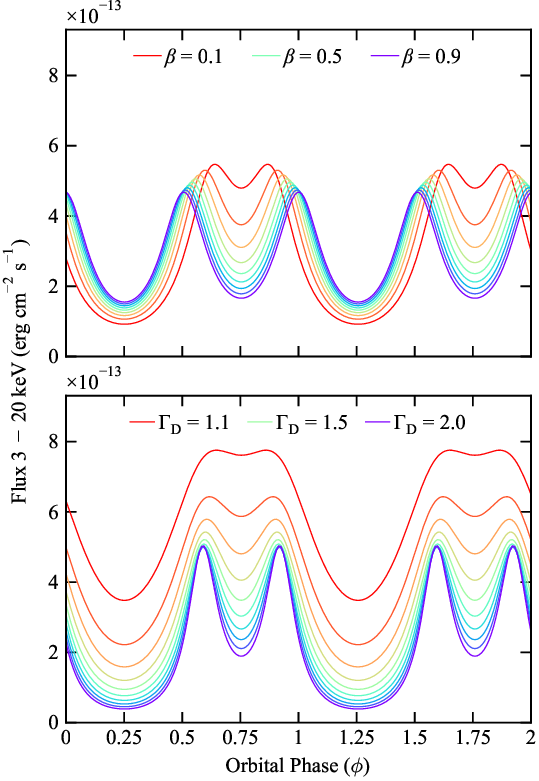}
\figcaption{X-ray LCs illustrating emissions from the IBS zone, computed with a range of
values for $\beta$ (top) and $\Gamma_{\rm D}$ (bottom).
The parameters for J2339 reported in Table~\ref{ta:ta4} serve as the baseline in this example.
\label{fig:fig9}
}
\vspace{0mm}
\end{figure}

We assume that an IBS is formed by the interaction between two isotropic winds from the pulsar and companion.
In this case, the IBS's shape is dictated by the momentum flux ratio
\begin{equation}
\label{eq:beta}
\beta=\frac{\dot E_{\rm SD}}{\dot M_* v_* c},
\end{equation}
where $\dot M_* v_*$ denotes the momentum flux of the companion wind \citep[][]{Canto1996}. 
Given specific values of $a_{\rm orb}$ and $\beta$ values (Tables~\ref{ta:ta3} and \ref{ta:ta4}),
the curve $r_{\rm IBS}(\theta_p)$ (pink in Figure~\ref{fig:fig7})
that defines the IBS in the cross-section plane
is computed using the formulas from \citet{Canto1996}:
\begin{equation}
\label{eq:IBSr}
r_{\rm IBS}=a_{\rm orb}\mathrm{sin}\theta_* \mathrm{csc}(\theta_p + \theta_*)
\end{equation}
and
\begin{equation}
\label{eq:IBStheta}
\theta_*\mathrm{cot}\theta_*=1+\beta(\theta_p\mathrm{cot}\theta_p -1).
\end{equation}
To construct the IBS surface, we rotate this $r_{\rm IBS}(\theta_p)$ curve around the line of centers.
Although $\dot E_{\rm SD}$ was well-measured for the pulsars in our target RBs, the companion's $\dot M_* v_*$
remains unknown. Therefore, we adjust $\beta$ to explain the observed SEDs and LCs
with the model. The IBS opening angle increases with increasing $\beta$, leading to a larger phase separation between
the LC peaks (Figure~\ref{fig:fig9}).

Another parameter that needs to be specified is the length of the IBS ($l_s$).
Energetic electrons, responsible for generating high-energy radiation, are predominantly situated in regions close to
the shock nose \citep[e.g.,][]{dlf15}. In this work, we adopt $l_s=4 a_{\rm orb}$.
However, we note that different values of $l_s$ can also explain the data well. The resulting LCs and SEDs
do not alter much, and any difference caused by changes in $l_s$ can be compensated for by adjusting other
parameters.

For the computation of the IBS emission, we discretize the IBS surface into $80\times180$ emission regions along
the $\theta_p$ and $\phi_p$ directions. This is necessary because both $B$ and the seed photon density ($u_*$) within the IBS
vary over its surface. The latter can be computed using the orbit and companion parameters (Table~\ref{ta:ta3}).
The geometry of the IC scattering, which produces anisotropy in the emission,
is computed at each of the $80\times180$ regions.
$B$ is modeled according to \citep[e.g.,][]{rs16}:
\begin{equation}
\label{eq:IBSB}
B(s)=B_s \left ( \frac{r_{\rm IBS}(s)}{R_0}\right )^{-1},
\end{equation}
where $B_s$ and $R_0$ denote $B$ and $r_p$ at the IBS nose ($s=0$), respectively.
The parameter $B_s$ influences the flux and is optimized during our modeling process.

Moreover, we consider the bulk motion of the electrons in the tangent direction of the IBS as follows.
We assume that the bulk motion of the electrons undergoes acceleration as they flow along the shock \citep[e.g.,][]{Bogovalov2008},
such that the bulk Lorentz factor of the flow is given by
\begin{equation}
\label{eq:bulkL}
\Gamma_{\rm IBS}(s)=1 + \frac{s}{l_s}(\Gamma_{\rm D} - 1).
\end{equation}

The energy distribution of the electrons within the IBS is modeled as a power law with
an index $p_1$ between
$\gamma_{e,\rm min}$ and $\gamma_{e,\rm max}$ in the flow rest frame.
In reality, the energy distribution is anticipated to vary across regions at different
distances ($s$; Figure~\ref{fig:fig7}) from the shock nose due to various effects,
including radiative cooling and bulk acceleration. For simplicity, we do not consider
the spectral changes between emission regions caused
by these effects; hence, the same spectral shape is applied across the IBS.
However, the energy budget is taken into consideration by comparing the particle energy flowing
out of the IBS (at its tail) with the pulsar's injection, as follows.

As electrons flow along the IBS in the $\theta_p$ direction, continuously replenished by the pulsar,
the number distribution along the IBS (integrated over $\phi_p$) is described by
continuity \citep[e.g.,][]{Canto1996} as
\begin{equation}
\label{eq:Ndist}
\dot N(\theta_p) = \frac{\dot N_w(1-\cos \theta_p(s))}{2}.
\end{equation}
Assuming a (normalized) power-law energy distribution of the electrons in the flow rest frame
\begin{equation}
\label{eq:PL}
\frac{dN}{d\gamma_e}=\frac{1-p_1}{\gamma_{e,\rm max}^{1-p_1} - \gamma_{e,\rm min}^{1-p_1}} \gamma_e^{-p_1},
\end{equation}
the particle energy in the observer frame is computed as
\begin{equation}
\label{eq:obsE}
E_p=\Gamma_{\rm D}\int_{\gamma_{s,\rm min}}^{\gamma_{s,\rm max}} \gamma_e m_e c^2 \frac{dN}{d\gamma_e} d\gamma_e.
\end{equation}
The number of particles exiting the emission zone per second is $\dot N(\theta_{p,\rm max}$), and
thus the energy budget of the pulsar is governed by
\begin{equation}
\label{eq:ebalance}
\dot N(\theta_{p,\rm max})E_p = \eta_s \dot E_{\rm SD} f_{\Omega},
\end{equation}
where $\theta_{p,\rm max}$ corresponds to $\theta_p$ at $l_{s,\rm max}$ (IBS tail).
Optimization of $\gamma_{e\rm,min}$, $\gamma_{e\rm,max}$, and $p_1$ is performed to explain
the observational (primarily X-ray) data while satisfying Equation~(\ref{eq:ebalance}).

The bulk flow induces emission anisotropy and orbital modulation in both synchrotron and IC emissions.
The strength of the emissions relies on the Doppler factor of the flow \citep[e.g.,][]{Dermer2009,ar17}:
\begin{equation}
\label{eq:bulkD}
\delta_{\rm IBS}=\frac{1}{\Gamma_{\rm IBS}(1-v_{\rm IBS}\cos\theta_V/c)},
\end{equation}
where $v_{\rm IBS}=c\sqrt{1 - 1/\Gamma_{\rm IBS}^2}$ and $\theta_V$ is the angle between the flow direction and the LoS.
Orbital changes in the LoS relative to the IBS (and the binary system) and thus $\theta_V$ induce the orbital modulation of the LCs.
For a given $i$ (Table~\ref{ta:ta3}) and the direction of flow (IBS tangent) at an orbital phase,
$\delta_{\rm IBS}$ in each of the $80\times180$ emission regions is determined by
only one parameter, $\Gamma_{\rm D}$. 
This parameter is optimized to explain the SEDs and LCs.
This optimization of $\Gamma_{\rm D}$ impacts the sharpness
of the LC peaks (e.g., see Figure~\ref{fig:fig9}) and
the number of electrons in the IBS (Equations~(\ref{eq:obsE}) and (\ref{eq:ebalance})).

The emission from the IBS remains essentially the same among the scenarios discussed in
this paper, with minor variation attributable to differences in the numbers of electrons injected
from the wind zone, induced by shock penetration (e.g., Equations~(\ref{eq:Et})--(\ref{eq:etatw})).
We should note that the IBS parameters are linked to the wind parameters (e.g., Equation~(\ref{eq:Ndist}));
these parameters are consistently adjusted.

\subsubsection{Companion zone}
\label{sec:sec4_1_3}
This zone starts at the IBS location ($r_p=r_{\rm IBS}$; Equation~(\ref{eq:IBSr})) and extends outward from the pulsar
(green region in Figure~\ref{fig:fig7}).
Given that the electrons in this zone are energetic and cold pulsar-wind particles
that have penetrated the IBS (Section~\ref{sec:sec3_2}),
they move ballistically away from the pulsar.
Similar to the wind zone (Section~\ref{sec:sec4_1_1}),
we consider emission solely from electrons traveling along the LoS (green arrow in Figure~\ref{fig:fig7}).
Hence, the emission zone essentially takes the form of a line.
We split this line into radial segments to account for varying conditions such as the companion's $B$,
assumed to have a dipole structure:
\begin{equation}
\label{eq:compb}
B(r_*)=B_c\left (\frac{r_*}{R_*} \right )^{-3}.
\end{equation}
The strong dependence of $B$ on $r_*$ implies a rapid decline in emission from
these electrons with increasing $r_*$.

\begin{figure}
\centering
\includegraphics[width=3.2 in]{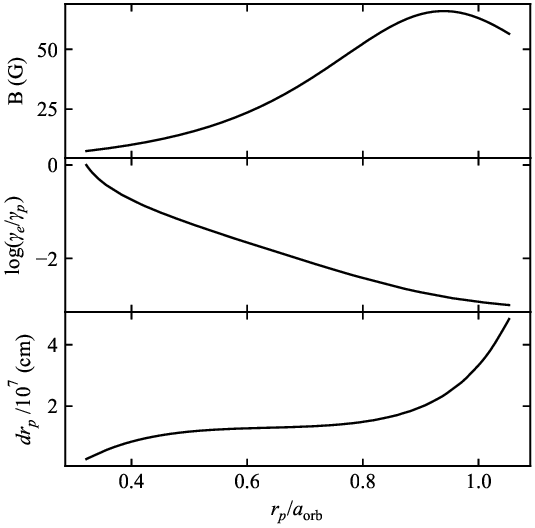}
\figcaption{$B$ (top), $\gamma_e$ (middle), and $dr_p$ (bottom) calculated for an emission region
within the companion zone in J2339 (Table~\ref{ta:ta4}) during the SUPC phase.
The emission region, characterized by a linear profile, initiates at $R_{\rm IBS}\approx0.3a_{\rm orb}$, where
electrons are energetic. In this phase, electrons traveling along the LoS  approach the companion at $a_{\rm orb}$.
Owing to the strong $B$, these electrons efficiently lose energy in the inner regions, and
the cooling timescale increases with $r_p$. The analysis involves 7000 radial bins to comprehensively
characterize the emission region at this phase.
\label{fig:fig10}
}
\vspace{0mm}
\end{figure}

We consider this zone only in Scenario~2.
In this scenario, we assume that the electrons injected by the pulsar
follow a $\delta$ distribution with a very high Lorentz factor ($\gamma_p\gapp 10^8$):
\begin{equation}
\label{eq:primary}
\frac{dN}{d\gamma_e dt}=\dot N_p\delta(\gamma_e-\gamma_p).
\end{equation}
While most of these unshocked primary electrons
in the wind zone transform into lower-energy electrons (secondaries), constituting
the ``wind zone'' (Section~\ref{sec:sec4_1_1}), a small
fraction ($\zeta$) of the primaries passes through the IBS
and emits in this companion zone. The energy budget of the pulsar
is governed by Equations~(\ref{eq:Et})--(\ref{eq:etatw}).

Due to our assumption of a very large value for $\gamma_p$ ($\ge 10^8$), the IC emission from
these shock-penetrating electrons is highly suppressed, whereas their
synchrotron emission under the strong $B$ of the companion is very intense.
The synchrotron cooling timescale (Equation~(\ref{eq:tcool})) of the electrons
is significantly shorter than their residence time, especially near the SUPC phase
where the electrons closely approach the companion (Figure~\ref{fig:fig4} right).
In the computation of the emission from this zone, we account for synchrotron cooling.
This necessitates the use of different lengths for the radial segments ($dr_p$; Figure~\ref{fig:fig10}),
ensuring that each segment is considerably shorter than $ct_{\rm cool}$ (Equation~\ref{eq:tcool}):
$dr_p=10^{-3}ct_{\rm cool}$. 

In the first segment (closest to the IBS), we inject
\begin{equation}
\label{eq:npseg}
\zeta \dot N_p\frac{dr_p}{c}
\end{equation}
electrons with a Lorentz factor of $\gamma_p$
and evolve them while accounting for their cooling (middle panel in Figure~\ref{fig:fig10}).
Similar to the approach for the wind zone, the solid-angle fraction $1/4\pi d^2$ is taken into account
in the emission formula (Equation~(\ref{eq:sysed})).
We integrate the emission up to a distance
where the Lorentz factor of the electrons drops to $10^{-3}\gamma_p$. If this length
proves to be excessively long, we terminate the integration at $l_{\rm comp}=4a_{\rm orb}$.
We verified that the emission from electrons beyond the integration region is negligibly small.

The emission SED and LC are primarily determined by $\zeta$, $\gamma_p$, and $B_c$ (Figure~\ref{fig:fig6}).
The combination of $\gamma_p$ and $B$ plays a crucial role in determining the frequency of the synchrotron SED peak
(e.g., Equation~(\ref{eq:synu})).
Achieving an appropriate balance in $B\gamma_p^2$ ensures that the peak emission occurs in the GeV band,
resulting in a high GeV flux.
Excessively high values of $\gamma_p$ or $B$ (e.g., near the SUPC phase) can push the SED peak beyond the GeV band,
leading to a reduction in the $\sim$GeV flux and causing a dip in the LC at some phases
($\phi=0.25$ in Figure~\ref{fig:fig6}).
On the contrary, if $B$ is low (away from the SUPC phase), the peak emission occurs
below the GeV band, resulting in a model-predicted GeV emission that is low (Figure~\ref{fig:fig5}).
Moreover, electrons do not radiate efficiently when $B$ is low.
We optimize these parameters to explain the observed GeV fluxes and LCs (Figure~\ref{fig:fig6}).

\subsection{Computation of the synchrotron and IC emissions}
\label{sec:sec4_2}
We calculate the synchrotron and IC emissions from electrons in each segment of the aforementioned zones
using the formulas detailed in \citet{Finke2008} and \citet{Dermer2009}.

\subsubsection{The synchrotron emission}
\label{sec:sec4_2_1}
The synchrotron SED from isotropically-distributed electrons (in the flow rest frame)
under the influence of a randomly-oriented $B$ is given in Equation~(18) of \citet{Finke2008}:
\begin{equation}
\label{eq:sysed}
f_{\rm SY}(\epsilon)=\frac{\sqrt{3}\delta_{\rm D}^4\epsilon'q_e^3B}{4\pi hd^2}\int_1^\infty d\gamma_e'
\frac{dN_e'(\gamma_e')}{d\gamma_e'}R(x),
\end{equation}
where primed quantities are defined in the flow rest frame.
In this formula, $\delta_{\rm D}$ represents the Doppler factor of the bulk flow (e.g., Equation~(\ref{eq:bulkD})),
$h$ is the Planck constant, and $q_e$ denotes the charge of an electron.
The observed and emitted photon energies are expressed in units of
$m_e c^2$ as $\epsilon$ ($\equiv h\nu/m_ec^2$) and $\epsilon'$ ($\equiv h\nu'/m_e c^2$), respectively.
$\frac{dN'_e(\gamma_e')}{d\gamma_e'}$ represents the energy distribution of the
emitting electrons (Equations~(\ref{eq:PL}) and (\ref{eq:primary})), and 
\begin{equation}
\label{eq:syrx}
R(x) = \frac{x}{2}\int_0^{\pi} d\theta \sin\theta \int_{x/\sin\theta}^{\infty} dt K_{5/3}(t),
\end{equation}
where $x\equiv \frac{4\pi \epsilon' m_e^2 c^3}{3q_e B h\gamma_e'^2}$, and
$K_{5/3}$ is the modified Bessel function of the third kind. For computations of $R(x)$, we use
approximate formulas provided by \citet{Finke2008}.
We carry out the computation of Equation~(\ref{eq:sysed}) using 700 bins for $\epsilon$
and 200 bins for $\gamma_e$.

While the assumptions underlying Equation~(\ref{eq:sysed}) are valid for the synchrotron
emission from the IBS, it is important to note that electrons in the companion zone are not isotropically distributed,
and $B$ of the companion is not randomly oriented.
Since the orientation of the companion's $B$ is unknown (we show an aligned case in Figures~\ref{fig:fig4} and \ref{fig:fig7} only for illustrative purposes), it is not possible to account for
pitch angle scattering. In this study, due to the lack of information about
the specific orientation, we take an average over the pitch
angle, as reflected in Equation~(\ref{eq:syrx}). 
The pitch angles experienced by shock-penetrating particles, and consequently, the amplitude and shape of the GeV LC,
will be influenced by the orientation of the companion's $B$. This aspect could be explored in a future paper.

\subsubsection{The IC emission}
\label{sec:sec4_2_2}
We use the IC emission formula given in \citet{Dermer2009}. This formula for the emission SED
incorporates various parameters, including Doppler boosting, anisotropic scattering, and
the seed photon spectrum, and is expressed as \citep[e.g., Equation~(34) of][]{Dermer2009}:
\begin{equation}
\label{eq:ecsed}
\begin{aligned}
f_{\rm IC}(\epsilon_s)= & \frac{3c\sigma_{\rm T}}{32\pi d^2}\epsilon_{\rm s}^2 \delta_{\rm D}^3\int_0^{2\pi}d\phi_*
\int_{-1}^1 d\mu_*
\int_0^{\epsilon_{*,\rm hi}} d\epsilon_* \times & \\
& \frac{u_*(\epsilon_*,\Omega_*)}{\epsilon_*^2}
\int_{\gamma_{\rm low}}^{\infty} d\gamma_e \frac{dN'_e(\gamma_e')}{d\gamma_e'}\frac{\Xi}{\gamma_e^2},&
\end{aligned}
\end{equation}
where $\gamma_e=\delta_{\rm D}\gamma_e'$.

As in the case of the synchrotron formula, the photon energies before ($\epsilon_*$)
and after ($\epsilon_s$) an IC scattering are normalized by $m_e c^2$.
The parameters $\phi_*$ and $\mu_*$ ($=\mathrm{cos}\theta_*$) define the direction of the seed photon, with the energy density of
$u_*(\epsilon_*, \Omega_*)$ in the emission zone. In this context, we consider only the companion's BB spectrum,
characterized by $R_*$ and $T_*$ (Table~\ref{ta:ta3}), as the source of the seed photons.
$\Xi$ is defined by
\begin{equation}
\label{eq:ecxi}
\Xi \equiv y + y^{-1} - \frac{2\epsilon_s}{\gamma_e \bar \epsilon y} +\left (\frac{\epsilon_s}{\gamma_e \bar \epsilon y} \right )^2,
\end{equation}
where $y\equiv 1-\frac{\epsilon_s}{\gamma_e}$ and
$\bar \epsilon$ represents the invariant collision energy
$\bar \epsilon \equiv \gamma_e \epsilon_*(1-\sqrt{1-1/\gamma_e^2}\mathrm{cos}\psi)$
with $\psi$ denoting the scattering angle.

The integration limits in Equation~(\ref{eq:ecsed}) are determined by the scattering kinematics:
\begin{equation}
\label{eq:glow}
\gamma_{\rm low} = \frac{\epsilon_s}{2}\left[ 1 + \sqrt{1 + \frac{2}{\epsilon_* \epsilon_s(1-\mathrm{cos}\psi)}} \right ]
\end{equation}
and
\begin{equation}
\label{eq:ehi}
\epsilon_{*,\rm hi}= \frac{2\epsilon_s}{1-\mathrm{cos}\psi}.
\end{equation}
We calculate the IC SED using 700 bins for $\epsilon_s$ and $\epsilon_*$,
and 200 bins for $\gamma_e$. In this study, we assume that the companion is a
point source, simplifying the $\phi_*$ and $\theta_*$ integrations in Equation~(\ref{eq:ecsed}).

In each emission region (segment) within the emission zones, as described in Section~\ref{sec:sec4_1},
we compute the necessary quantities for emissions,
such as $\delta_{\rm D}$, $B(s)$, $B(r_*)$, $u_*(\epsilon_*,\Omega_*$), and $\psi$,
taking into account the orbit (e.g., direction of the LoS) and the geometry of the emission region
(e.g., $r_*$, $r_p$ and the flow direction).
We combine these with the particle distribution $dN/d\gamma_e$ within the region to compute the emission SEDs
at 50 orbital phases.
Note that the energy distributions of the mono-energetic electrons in the wind (Equation~(\ref{eq:deltaenergy}))
and companion zones (Equation~(\ref{eq:primary})) are given
in the observer rest frame, and thus their emissions are not Doppler boosted (i.e., $\delta_{\rm D}=1$).
Conversely, the distributions of electrons in
the decelerated wind (Equation~(\ref{eq:Maxwell})) and IBS (Equation~(\ref{eq:PL})) zones
were expressed in the flow rest frame (these electrons have bulk flow speeds), and
their emissions are Doppler boosted.
The synchrotron X-rays exhibit orbital modulation due to this Doppler boosting,
while the GeV modulation results from changes in either the IC scattering geometry
(e.g., $\psi$; scenarios~1 and 1a) or the cooling timescale determined by the companion's $B(r_*)$ (Scenario~2).

\begin{table}
\begin{center}
\caption{Parameters used for models displayed in Figures~\ref{fig:fig1} and \ref{fig:fig3}}
\label{ta:ta4}
\scriptsize{
\begin{tabular}{lcccc}
\hline \hline
Property		&	Unit	&	J1227	&	J2039	&	J2339  \\  \hline
\multicolumn{5}{l}{Common to Scenarios~1 and 2} \\
$p_1$			&		&	1.56	&	1.34	&	1.37  \\ 
$\gamma_{s,\rm min}$	&		&	16	&	1	&	1  \\ 
$\gamma_{s,\rm max}$	&	$10^6$	&	10	&	6	&	6  \\ 
$\gamma_w$		&	$10^4$	&	0.82	&	2.14	&	1.61  \\ 
$\Gamma_{\rm D}$	&		&	1.15	&	1.37	&	1.45  \\ 
$\beta$			&		&	0.20	&	0.24	&	0.20  \\  \hline
\multicolumn{5}{l}{Scenario~1} \\
$B_s$			&	G	&	1.93	&	3.30	&	3.80   \\ 
$\eta_w$		&		&	0.95	&	0.79	&	0.82 \\ 
$\eta_s$		&		&	0.95	&	0.79	&	0.82 \\  \hline
\multicolumn{5}{l}{Scenario 2} \\
$B_s$			&	G	&	1.97	&	3.70	&	4.10   \\ 
$\eta_w$		&		&	0.92	&	0.67	&	0.73 \\ 
$\eta_s$		&		&	0.92	&	0.67	&	0.73 \\ 
$B_c$			&	kG	&	2.35	&	1.52	&	1.39   \\ 
$\eta_p$		&		&	0.95	&	0.79	&	0.82 \\
$\zeta$			&		&	0.03	&	0.15	&	0.11  \\ 
$\gamma_p$		&	$10^7$	&	7	&	20	&	4  \\
$\mathcal{M}$		&	$10^3$	&	8.6	&	9.5	&	2.5  \\ \hline
\end{tabular}}
\end{center}
\end{table}

\subsection{Results of modeling}
\label{sec:sec4_3}
We initiated our model with the rough estimations provided in Section~\ref{sec:sec3} as the starting
point for the parameters. Subsequently, we fine-tuned these parameters to match the observed
SEDs and LCs of the RB targets with the model computations. The optimized parameter values are outlined in Table~\ref{ta:ta4}.
Due to the large number of parameters, a formal data fitting process was challenging.
Instead, we employed a qualitative `fit-by-eye' approach.
Therefore, we do not report uncertainties in the estimated parameter values;
we defer this to future work.

Figures~\ref{fig:fig1} and \ref{fig:fig3} display the computed SEDs and LCs corresponding to the three cases
described earlier (see Section~\ref{sec:sec3_2}):
the basic scenario (red; Scenario~1), its modification (blue; Scenario~1a), and
Scenario~2 (green).
The predicted GeV fluxes and LCs exhibit significant variations among the scenarios.
Similar to Scenario\,1, the GeV LCs in Scenario\,1a (Figure~\ref{fig:fig3}) exhibit orbital
modulation due to the changing IC scattering geometry. However, they are slightly broader
than the LCs predicted by Scenario\,1 due to the isotropization and thermalization of the electrons.
Scenario~1 is unable to explain the LAT data,
whereas Scenario~2 can readily explain the multiband data.

\begin{figure*}
\centering
\includegraphics[width=7 in]{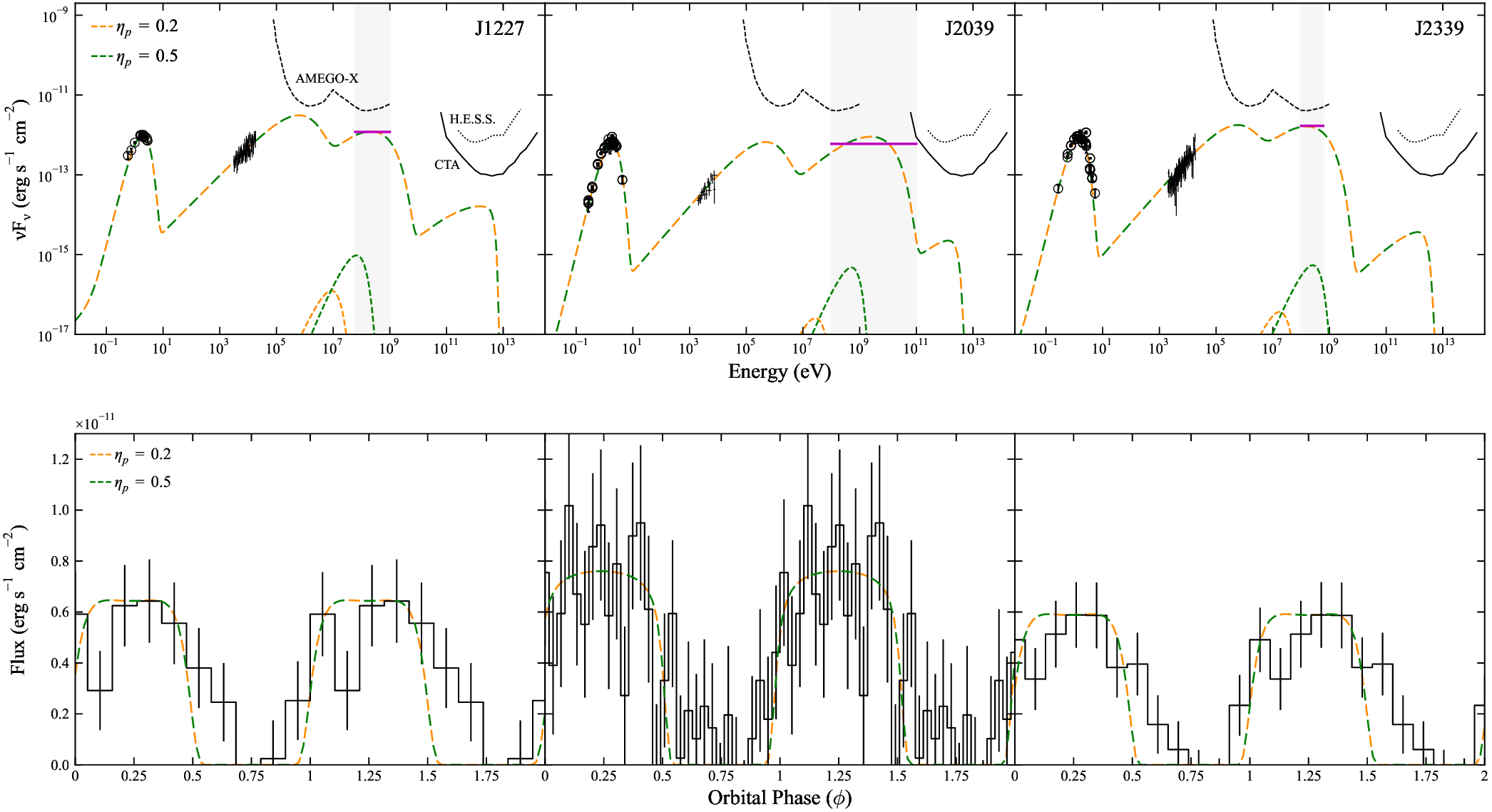}
\figcaption{Model-generated SEDs and LCs are depicted for various $\eta_p$ values for Scenario~2.
The data points are the same as those in Figure~\ref{fig:fig3}.
The orange and green curves correspond to models with $\eta_p=$0.2 and 0.5, respectively.
In this example, the decrease in $\eta_p$ (compared to our baseline value in Table~\ref{ta:ta4})
was compensated for by an increase in $\zeta$.
\label{fig:fig11}
}
\vspace{0mm}
\end{figure*}

The parameter values derived from the basic model (Scenario~1) are in accord with those
obtained using the analytic approach (Section~\ref{sec:sec3_2}). To match
the X-ray SED, the IBS electrons should follow a power-law distribution
with an index of $p_1\approx $1.3--1.6
between $\gamma_{s,\rm min}$ and $\gamma_{s,\rm max}$.
These two parameters were adjusted to explain the X-ray data and not to overpredict
the LAT flux for the given $\eta_s$ value (Table~\ref{ta:ta3})
of each target.
The highest-energy electrons in the IBS upscatter the stellar BB emission to TeV energies.
Therefore, we relied on the IC emission from the wind zone to generate the peak of the IC SED at $\lapp$GeV.
As a result, the basic model (red curves in the top row of Figure~\ref{fig:fig3})
exhibits two peaks in the $>$MeV SED.
The low-energy peak, centered around $\gamma_w^2 h\nu_*\sim$100\,MeV,
corresponds to the IC emission from the wind zone, while the
high-energy peak, around $\gamma_{s,\rm max}^2 h\nu_*\sim$TeV, corresponds to
the IBS IC emission \citep[see][for further discussion on this component]{Merwe+2020,Wadiasingh2022}.
While the computed LC shapes of Scenario~1 resemble the observed ones, the predicted gamma-ray
fluxes were found to be insufficient, as mentioned in Section~\ref{sec:sec3_2}.

The lack of GeV fluxes in this scenario was addressed by increasing the residence time
of the electrons within the wind zone, achieved through flow
deceleration (Scenario~1a). Despite our analytic investigation suggesting a very low value of $v_{\rm wind}$
(Equation~(\ref{eq:fluxratio})), we sought a validation using
our numerical model. The model with a constant speed profile
resulted in wind speeds (500\,\kms--5000\,\kms) higher than our rough estimations in Section~\ref{sec:sec3_2_1},
yet still too low to induce shock formation.
Exploring different speed profiles, such as linear or exponential decreases in $v_{\rm wind}$
with increasing distance from the light cylinder (toward the IBS),
cannot provide a remedy as these alternatives would require even lower values of $v_{\rm wind}$ at the position
of the wind-wind interaction region (i.e., IBS). $B$ energy in the wind zone
may be substantial \citep[][]{Cortes2022}, leading to a smaller $\eta_w$ than the assumed $\eta_w=\eta_s$ (Table~\ref{ta:ta4}).
In this case, further reductions in $v_{\rm wind}$ compound the challenges associated with this scenario.

Scenario~2 can reproduce the multiband data
with a companion's surface $B_c$ of $\gapp$\,kG, which 
falls within the range suggested based on physical arguments \citep[][]{Wadiasingh2018,Conrad2023}.
The gamma-ray SEDs predicted by this scenario are broad due to the effects
of energy loss and variation in $B$. The predicted GeV LCs exhibit a flat top
with a rapid decline at the phases corresponding to the IBS tangent
(see Figure~\ref{fig:fig5} and Section~\ref{sec:sec4_1}).
The parameter values for the IBS are very similar to those used for Scenario~1,
as $\zeta$ is not very high. Due to the loss of particles,
the IBS synchrotron emission (X-ray) was slightly lower.
This was addressed by using a stronger $B_s$ (Table~\ref{ta:ta4}),
but it could also be achieved by a smaller $\gamma_w$ equally well.

It is important to note that the model parameters are highly covariant,
meaning that the values presented in Table~\ref{ta:ta4} are
not unique (see Section~\ref{sec:sec5}). In this work, we used the maximum possible
values for $\eta_p$ (i.e., $=1-\eta_\gamma$) and $\eta_w$ for Scenario~2. A lower value of $\eta_p$
\citep[e.g., see][]{Cortes2022}
would weaken the GeV emission, and this could be compensated for by increasing $\zeta$
and/or $B_c$ (Figure~\ref{fig:fig11}).

\section{Discussion and Summary}
\label{sec:sec5}
We analyzed the X-ray observations of three RB pulsars from which orbitally-modulating
GeV signals were detected. We then constructed their multiband SEDs and LCs
and investigated potential scenarios for the gamma-ray modulations using a phenomenological IBS model.

Based on our modeling, we found that Scenario~1 is unable to explain the measured GeV
fluxes of the RB targets,
as previously noted \citep[][]{ark20, Clark+2021}.
It is worth noting that our computations might underestimate the GeV flux since
electrons in regions with higher inclinations, such as those near the orbital plane,
can see stronger BB emission (from the heated surface of the companion) than what we assumed
(i.e., observed at an inclination angle $i<90^\circ$).
On average, however, electrons spread over extended emission zones
(Figure~\ref{fig:fig4}) do not preferentially encounter the most intense photon field,
and so the increase in the GeV flux due to this effect would be modest.
A further increase in the IC flux can be achieved if the distances to the sources
were smaller and the pulsar's $\dot E_{\rm SD}$
is larger (Equation~(\ref{eq:Efrac})). The latter may be possible since
neutron stars in pulsar binaries may be more massive
\citep[e.g., $\gapp2M_\odot$;][]{sh14,Romani2022} than $1.4 M_\odot$
used for the $\dot E_{\rm SD}$ estimation.
These increases are not very large and thus would be still insufficient to explain the
orders-of-magnitude discrepancy in the GeV band (red in Figure~\ref{fig:fig3}).
As demonstrated in the previous sections, addressing this issue involves a bulk deceleration of the unshocked wind to a low speed. However, this approach lacks physical support.
If the wind zone does not account for gamma-ray production,
there must be an alternative physical process capable of retaining the
$\gamma_w\approx 10^4$ electrons within the system for an extended period
for this scenario to be plausible.

Alternatively, the IBS model constructed based on Scenario~2
could easily accommodate the broadband SEDs and multiband LCs of the three RBs (Figure~\ref{fig:fig3}).
However, we should note that the parameter values reported in Table~\ref{ta:ta4}
are not unique due to parameter degeneracies \citep[see][]{KimAn2022}, and so MeV and
TeV flux predictions here should be taken as one potential realization among a landscape of possibilities.
Some of the degeneracy can be broken by high-quality optical, X-ray, MeV and TeV data or limits.
$i$ and $\eta_s$ values can be inferred from optical LC modeling \citep[e.g.,][]{Breton2013} and
the LAT measurement of the pulsar flux, respectively. The X-ray data help constrain
$v_{\rm IBS}$ \citep[by widths and amplitudes of the LC peaks;][]{KimAn2022},
$B_s$ within the IBS (Equation~(\ref{eq:B})), and $\gamma_{s,\rm max}$ (synchrotron cut-off).
Current constraints on these parameters are not stringent because
the quality of the X-ray LC measurements is poor and the SEDs do
not fully cover the energy range of the synchrotron spectrum.
For the $B_s$ values in Table~\ref{ta:ta4}, our model predicts a
synchrotron spectral cut-off at $\ge$100\,keV (Figure~\ref{fig:fig3}),
which can be confirmed by future hard X-ray and/or soft gamma-ray observations.

The observed GeV LCs appear broader than
those predicted by Scenario~2 (green in Figure~\ref{fig:fig3} bottom).
This might be caused by the observational effect that the probability-weighted LAT LCs show
reduced modulation compared to the sources' actual variability \citep[e.g.,][]{Corbet2016},
meaning that the intrinsic GeV LCs of the targets may be
narrower than the observed ones. Moreover, the LC shape may also alter
depending on the flat levels (constant emission) that are subtracted from the LCs.
In addition, Scenario~2 predicts broader SEDs in the GeV band, compared,
for example, with the J2039 SED obtained from differencing LAT SEDs of
the orbital maximum and minimum intervals \citep[][]{Clark+2021}.
A more direct measurement of the SED of the orbitally-modulating GeV emission,
facilitated by the LAT and/or other GeV observatories \citep[e.g., AMEGO-X;][]{Fleischhack2022},
will contribute to advancing our studies based on Scenario~2.

It was suggested that the `pulsed flux'
is orbitally modulated in J2039 and J2339 \citep[][]{Clark+2021, ark20}, implying
that the pulsar's pulsations should be preserved in the
emission regions of the modulating gamma-ray signals.
To maintain the pulse structure, the flow in the emission zone should
have a striped-wind structure, and the emission timescale (cooling timescale)
should be considerably shorter than the pulse periods of the pulsars.
The latter condition is satisfied by Scenario~2, in which
the cooling timescale is shorter than a millisecond (Equation~(\ref{eq:tcool})).
If the flow structure in the companion zone preserves the striped structure,
Scenario~2 can potentially explain the orbital modulation of the pulsed signals as well.
However, the striped wind might be destroyed in the wind zone or in the IBS;
this requires further theoretical studies.

As demonstrated earlier, Scenario~2 exhibits favorable aspects in explaining the observed data for the
RB targets. While
further theoretical studies with physically-motivated models for pulsars and their winds
\citep[e.g.,][]{Sironi2011,Cortes2022,Kalapotharakos2023} and high-quality measurements
are needed to confirm the scenario,
it is almost certain that the unshocked wind plays an important role in pulsar binaries.
Scenario~2 requires high-energy particles attaining $\sim$PeV, close to the voltage drop available to millisecond pulsars.
This would mean that pulsar magnetospheres are very efficient particle accelerators, with implications
for pulsed (magnetospheric) TeV emission from pulsars \citep[][]{Harding2021,HESS2023Vela}.
This is also in accord with some acceleration and emission scenarios for pulsed GeV emission
from pulsars (i.e. the curvature radiation scenario).

An important factor that we did not consider in this study
is the potential synchrotron emission which may originate from the
`current sheets' within the striped wind.
In regions near the pulsar's light cylinder where $B$ is expected to be strong,
the synchrotron emission would fall within the GeV band and could
add to the pulsed flux of the pulsar \citep[][]{Petri2012b}.
On the other hand, in regions near the IBS, we speculate that
$B$ is low \citep[e.g., $\approx B_s/3<1$\,G;][]{kc84a}
due to the dissipation of magnetic energy in the current sheets \citep[][]{Sironi2011}.
In this case, the synchrotron emission frequency would be $\le 1$\,eV
(Equation~(\ref{eq:synu})). An important question is
``how much of the magnetic energy is converted to particles and to radiation in the wind zone?''
This may have a significant impact on the structure of and emission from the
pulsar wind. Accurate measurements of the broadband SEDs and LCs of pulsar binaries, and
further PIC simulations for pulsar winds (including radiation) are warranted.

We adopted the simplicity of the IBS shape for a two isotropic wind interaction \citep[Section~\ref{sec:sec4_2_1};][]{Canto1996}. The single parameter ($\beta$) allows for considering a range of opening angles, but it may not capture the complexities of anisotropic winds or wind-magnetosphere interactions. In these more intricate cases, the IBS shape can exhibit greater complexity, influenced by system parameters such as the pulsar's spin axis orientation or the companion's magnetic field \citep[][]{Wadiasingh2018, kra19}. While the isotropic wind model reproduces the general features of the IBS in the other cases, subtle discrepancies may arise in the predicted LCs. The wind-$B$ interaction model offers the additional benefit of determining the $B$ orientation encountered by shock-penetrating electrons. Further investigations based on the wind-$B$ interaction model hold the potential to unlock new insights into the gamma-ray emission mechanisms of RBs.

Although our simplified IBS model with phenomenological prescriptions for the wind flows
may not encompass all the important physics of the flows,
we demonstrated that Scenario~2 offers a
plausible explanation for the orbitally-modulating GeV signals detected
from a few RBs. This scenario, along with our IBS model, can be further tested
with high-quality X-ray and gamma-ray data, and provide an opportunity
to comprehend the energy conversion and particle flow within the pulsar wind zone.
Future X-ray and gamma-ray observatories such as AXIS \citep[][]{Reynolds2023},
HEX-P \citep[][]{Madsen2023}, and AMEGO-X \citep[][]{Fleischhack2022},
have potential to furnish high-quality data for more samples, thereby facilitating
better understanding of the IBS and pulsar wind.

\bigskip

\begin{acknowledgments}
This work used data from the NuSTAR mission, a project led by the California Institute of Technology,
managed by the Jet Propulsion Laboratory, and funded by NASA. We made use of the NuSTAR Data
Analysis Software (NuSTARDAS) jointly developed by the ASI Science Data Center (ASDC, Italy)
and the California Institute of Technology (USA).
Z.W. thanks Jeremy Hare for interesting discussions.
Z.W. acknowledges support by NASA under award number 80GSFC21M0002. 
This work was supported by the National Research Foundation of Korea (NRF)
grant funded by the Korean Governmaent (MSIT) (NRF-2023R1A2C1002718).
This paper employs a list of Chandra datasets, obtained by the Chandra X-ray Observatory,
contained in~\dataset[DOI:10.25574]{https://doi.org/10.25574/cdc.165}.
We thank the referee for the insightful comments that helped improve the clarity of the paper.
\end{acknowledgments}

\vspace{5mm}
\facilities{CXO, XMM-Newton, NuSTAR}
\software{HEAsoft \citep[v6.31.1;][]{heasarc2014}, CIAO \citep[v4.12;][]{CIAO2013},
XMM-SAS \citep[20230412\_1735;][]{xmmsas14}, XSPEC \citep[v12.13.0c;][]{a96}}

\bibliographystyle{apj}
\bibliography{msrev2.bbl}

\end{document}